\documentclass[%
 reprint,
superscriptaddress,
 amsmath,amssymb,
 aps,
]{revtex4-2}

\usepackage{graphicx}
\usepackage{dcolumn}
\usepackage{bm}
\usepackage{upgreek}

\newcommand{\kbt}{k_\mathrm{B}T}
\newcommand{\microm}{\upmu\textrm{m}}

\begin{document}

\preprint{APS/123-QED}

\title{Erythrocyte sedimentation:\\ Effect of aggregation energy on gel structure during collapse}

\author{Anil Kumar Dasanna}
\affiliation{Theoretical Physics of Living Matter, Institute of Biological Information Processing and Institute for Advanced Simulation, Forschungszentrum J\"ulich, 52425 J\"ulich, Germany}
\author{Alexis Darras}
\email{alexis.charles.darras@gmail.com}
\affiliation{Experimental Physics, Saarland University, 66123 Saarbruecken, Germany
}
\author{Thomas John}%
\affiliation{Experimental Physics, Saarland University, 66123 Saarbruecken, Germany
}
\author{Gerhard Gompper}
\affiliation{Theoretical Physics of Living Matter, Institute of Biological Information Processing and Institute for Advanced Simulation, Forschungszentrum J\"ulich, 52425 J\"ulich, Germany}
\author{Lars Kaestner}
\affiliation{Experimental Physics, Saarland University, 66123 Saarbruecken, Germany
}
\affiliation{Theoretical Medicine and Biosciences, Saarland University, 66424 Homburg, Germany}
\author{Christian Wagner}
\affiliation{Experimental Physics, Saarland University, 66123 Saarbruecken, Germany
}
\affiliation{Physics and Materials Science Research Unit, University of Luxembourg, Luxembourg City, Luxembourg}
\author{Dmitry A. Fedosov}
\affiliation{Theoretical Physics of Living Matter, Institute of Biological Information Processing and Institute for Advanced Simulation, Forschungszentrum J\"ulich, 52425 J\"ulich, Germany}

\date{\today}
\begin{abstract}
The erythrocyte (or red blood cell) sedimentation rate (ESR) is commonly interpreted as a measure of cell aggregation and as a biomarker of inflammation. It is  well known that an increase of fibrinogen concentration, an aggregation-inducing protein for erythrocytes, leads to an increase of the sedimentation rate of erythrocytes, which is generally explained through the formation and faster settling of large disjoint aggregates. However, many aspects of erythrocyte sedimentation conform well with the collapse of a colloidal gel rather than with the sedimentation of disjoint aggregates. Using experiments and cell-level numerical simulations, we systematically investigate the dependence of ESR on fibrinogen concentration and its relation to the microstructure of the gel-like erythrocyte suspension. We show that for physiological aggregation interactions, an increase in the attraction strength between cells results in a cell network with larger void spaces. This geometrical change in the network structure occurs due to anisotropic shape and deformability of erythrocytes and leads to an increased gel permeability and faster sedimentation. Our results provide a comprehensive relation between the ESR and the cell-level structure of erythrocyte suspensions and support the gel hypothesis in the interpretation of blood sedimentation.      
\end{abstract}

\maketitle

\section{Introduction}

The erythrocyte sedimentation rate (ESR)  is a common blood test that measures how fast red blood cells (erythrocytes) settle at the bottom of a test tube that contains an anti-coagulated but undiluted blood sample, composed of cells and blood plasma as carrier liquid. It is one of the oldest nonspecific medical tests for inflammation, and is still a gold standard for diagnosis and monitoring of inflammatory diseases \cite{lapic2020erythrocyte,lapic2020Corona}. Its first description dates back to the 19th century, but a similar procedure was already in use by ancient Greeks \cite{mccabe1985brief}.

Interestingly, the sedimentation rate is usually much faster than one would expect from the sedimentation of a single cells in a highly diluted sample. A common physical interpretation of the ESR is that erythrocytes suspended in blood plasma sediment as separate large aggregates, which is amenable to a theoretical description using the Stokes law for the drag force \cite{taye2020sedimentation,baskurt2011red,smallwood1985physics,puccini1977erythrocyte,dorrington1983erythrocyte,sousa2018validation}. However, the state of the art in colloidal science indicates that particle suspensions with a high volume fraction of weakly attractive colloids form a percolating network, which is generally referred to as a soft gel \cite{rouwhorst2020nonequilibrium,guo2011gel,teece2014gels,buscall2009towards,gopalakrishnan2006linking,padmanabhan2018gravitational,bartlett2012sudden,manley2005gravitational,derec2003rapid,allain1995aggregation}. 

\begin{figure}[ht!b]
\includegraphics[width=0.45\textwidth]{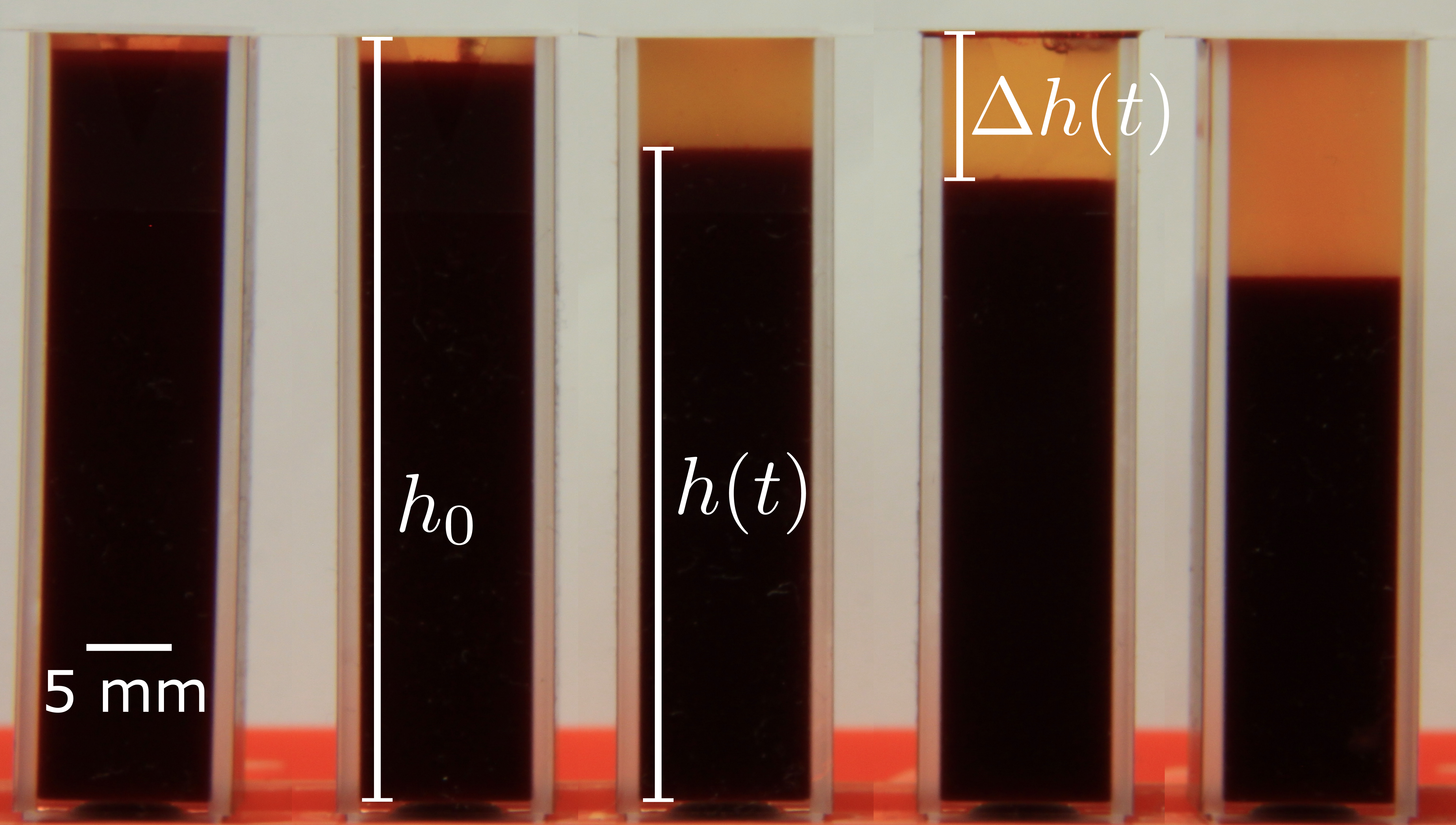}
\caption{\label{Illus} ESR experiments. Cuvettes containing blood samples with various levels of fibrinogen are shown after $2$ hours at rest. Hematocrit (i.e. erythrocyte volume fraction) in all containers has been adjusted to $\phi=0.45$. The very left container corresponds to erythrocytes suspended in autologous serum (no fibrinogen), while the very right contains erythrocytes in autologous blood plasma (maximum amount of fibrinogen). Middle containers contain cells suspended in a mixture of serum and plasma, with volume proportions of $25\%$, $50\%$ and $75\%$ of plasma, from left to right. The various height characteristics ($h_0$, $h(t)$ and $\Delta h(t)$) are also indicated.}
\end{figure}

In the joint letter \cite{JointLetter}, we have demonstrated that sedimenting erythrocytes at high volume fractions indeed also form a gel, which explains pertinent features of the sedimentation process and leads to a consistent theoretical description of sedimentation dynamics. An interesting observation for blood sedimentation is that the ESR increases with increasing the attraction between suspended cells \cite{baskurt2011red,pribush2010mechanism1}. This is exactly opposite to the behavior of hard-colloid suspensions, where an increase in attractive interactions between colloids results in gel stabilization, which can significantly delay and slow down the sedimentation process \cite{manley2005gravitational,channell2000effects,kilfoil2003dynamics,kamp2009universal,dinsmore2006microscopic,pribush2010mechanism1,teece2014gels,lindstrom2012}. This observation is of particular importance for medical applications which generally use the ESR as an inflammation marker. The current rationale for this marker is that inflammation correlates with high fibrinogen levels, which enhances the aggregation of erythrocytes. Therefore, a faster ESR for stronger aggregation interactions is believed to be related to the aggregation-mediated formation of large separate aggregates, whose sedimentation is faster in the Stokes regime \cite{bedell1985,kratz2017}. The dependence of ESR on fibrinogen concentration is illustrated in Fig. \ref{Illus}.

To systematically investigate the behavior of ESR on fibrinogen concentration and its relation to the micro-structure of erythrocyte suspension, we employ a combination of experiments and numerical simulations. We show that, in contrast to classical hard-colloid suspensions \cite{manley2005gravitational,allain1995aggregation,derec2003rapid}, an increase in attraction strength between cells significantly modifies the structure and permeability of erythrocyte gel, resulting in a faster ESR. This sedimentation behavior is due to the characteristic discocyte shape of erythrocytes and their membrane flexibility. The discocyte geometry facilitates the formation of well-known rouleaux stacks, which are eventually arranged into a network-like structure. More importantly, the deformability of RBCs implies an increase in ESR at high fibrinogen concentrations, as simulations demonstrate that the dependence of ESR on the attraction strength nearly disappears for stiff erythrocytes. These results provide a consistent explanation for the behavior of ESR as a function of fibrinogen concentration, and further support the gel hypothesis for erythrocyte suspensions.

The paper is organized as follows. Section \ref{ThRecal} summarizes the theoretical model for gel permeability and dynamic sedimentation introduced in the joint letter \cite{JointLetter}, since changes in the associated model parameters form a basis for the analysis of presented data. Experimental and numerical methods are given in Section \ref{Methods}. Section \ref{Results} presents the dependence of sedimentation on fibrinogen concentration from experimental measurements, along its correlation with the micro-structure of erythrocyte suspensions.

\section{Theoretical Description of Sedimentation}
\label{ThRecal}

We start with a brief summary of the theoretical model for gel structure and permeability developed in the joint letter \cite{JointLetter}, since model parameters are frequently used in data analysis. To describe the gravitational collapse of erythrocyte gel, a time-dependent volume fraction $\phi (t)$ is considered, covering the range from dilute to close-packed cell suspensions. For simplicity, we assume that only gravitational forces drive gel sedimentation, and a Carman-Kozeny relationship with the maximal volume fraction $\phi_m$ smaller than unity is used to approximate gel permeability \cite{terzaghi1925,carman1939permeability,ozgumus2014determination,heijs1995numerical,xu2008developing}. The characteristic velocity for the gravitational collapse of an erythrocyte suspension is obtained on the basis of Darcy's law for the flow of a fluid through a porous medium as
\begin{equation}
    -\frac{dh}{dt}=\frac{\Delta \rho g a^2}{\kappa_0 \eta}\frac{\left(\phi_m-\phi\right)^3}{\phi\left(1-\phi\right)},
    \label{VelOne}
\end{equation}
where $h(t)$ is the height of the gel at time $t$, $\Delta \rho$ is the density difference between erythrocytes and blood plasma, $g$ is the gravitational acceleration, $\eta$ is the viscosity of the suspending liquid, $a$ is the characteristic size of pores within the gel structure, and $\kappa_0$ is the dimensionless scaling constant of the Carman-Kozeny relationship. Here, the conservation of the volume of erythrocytes provides the necessary relation between the height $h$ and the volume fraction $\phi$ via $h(t) \phi(t)  = h_0 \phi_0$, where $h_0$ and $\phi_0$ are the initial height and volume fraction at $t=0$, respectively. A typical size of $a=R r_{\text{RBC}}$ is considered, where $r_{\text{RBC}}=4\,$\textmu{m} is the average radius of an erythrocyte and $R$ is a parameter reflecting the ratio between the size of erythrocytes and a characteristic size of fluid channels within the gel \cite{derec2003rapid,allain1995aggregation,teece2014gels}. The density difference is approximately $\Delta \rho\approx80\,\mathrm{kg/m^3}$ \cite{norouzi2017sorting,trudnowski1974specific}. Starting from the initial conditions for the height $h_0$ and volume fraction $\phi_0$, and taking into account the usual delay time $t_0$ for the gel collapse \cite{gopalakrishnan2006linking,teece2014gels,lindstrom2012}, the height of the gel as a function of time becomes
\begin{eqnarray}
    h(t)=
    \begin{cases}
    h_0,&\quad\text{if}~t<t_0\\
    f^{-1}\left(f(h_0)-\frac{G \phi_m^3}{\gamma} (t-t_0)\right)&\quad\text{if}~t\geq t_0
    \end{cases},
    \label{FinalEq}
\end{eqnarray}
where $G=\Delta \rho g r_{\text{RBC}}^2/(\eta\phi_0 h_0)$ is proportional to the inverse characteristic time for the sedimentation of single cells, and $\gamma=\kappa_0/R^2$ is a dimensionless fit parameter related to a characteristic time of the system. The third parameter $\phi_m$ of this model is also included in the function $f$, obtained as
\begin{align}
    f(x)&=\log{\left(\phi_m x -\phi_0 h_0\right)} +\nonumber\\&\frac{\phi_0 h_0\bigl(\phi_0 h_0 (3-\phi_m)-2(2-\phi_m)\phi_m x\bigr)}{2\left(\phi_0 h_0 -\phi_m x\right)^2}.
    \label{DetailMain}
\end{align}

It is instructive to consider the asymptotic behavior of the time dependence of $h(t)$ for small and large volume fractions, corresponding to short and large times, respectively. An expansion of right-hand side of Eq.\,\ref{VelOne} for small $\phi$ yields the behavior
\begin{equation}
    \Delta h = h_0 - h(t) \approx h_0 \left[ 1-\exp(-(t-t_0)/t^* \right]
\end{equation}
with the characteristic time scale $t^*=\gamma / \left(G \phi_m^3\right)$, while small $\delta \phi= \phi - \phi_m$ yields to
\begin{equation}
    \delta h(t) = h(t) - h_m \approx h_m \left(\frac{t}{t^*}\right)^{-1/2} \sqrt{\frac{1-\phi_m}{2}}
\end{equation}
where $h_m$ is the height corresponding to $\phi_m$.

The comparison of erythrocyte sedimentation under different conditions includes the three fitting parameters: the maximum volume fraction $\phi_m$, the dimensionless characteristic collapse time $\gamma$, and the delay time $t_0$ of the gel. The dependence of these parameters on the fibrinogen concentration, which modifies the aggregation strength between erythrocytes, will be presented, and the behavior of $\gamma$ with respect to the aggregation strength will be discussed.

\section{Methods}
\label{Methods}

\subsection{Macroscopic Measurements of Sedimentation}

Blood samples were collected from healthy volunteers after an informed consent, in compliance with the declaration of Helsinki, as approved by the 'Ärztekammer des Saarlandes', ethics votum 51/18.

The sedimentation process is monitored by taking an image sequence of a translucent cuvette containing blood with a frame rate of one frame per minute (see Fig.~\ref{Illus} for an example of the pictures). The cuvette has an inner square cross-section of $10\,\mathrm{mm}\times10\,\mathrm{mm}$, and a height of $40\,\mathrm{mm}$. In all investigations, the interface is sharp. The height of the packed erythrocyte phase is determined using a customly written Matlab code, as follows. The image is first binarized with the Otsu threshold, using the difference between the red and the blue channels as intensity. The binarized image is then averaged horizontally to obtain an intensity profile with minimal noise. The position of the interface is then determined as the position of the maximum intensity gradient. The pixel resolution leads to a spatial accuracy of approximately $0.1\,\mathrm{mm}$.

Sedimentation recordings are performed for different levels of fibrinogen, in order to quantify its effect on the parameters of the theoretical model. We focus on fibrinogen because it is one of the most efficient aggregation-promoting agents of erythrocytes \cite{brust2014plasma,flormann2015there}. The aggregation energy between erythrocytes, defined as the work required to separate two aggregated cells, is known to increase linearly with fibrinogen concentration for a wide range of concentrations, in contrast to some other aggregation-inducing molecules such as Dextran \cite{brust2014plasma,lee2016optical}. The initial volume fraction of cells was always kept at $\phi_0=0.45$, independently of fibrinogen concentration. To closely approximate physiological conditions, red blood cells (RBCs) are suspended into well-controlled mixtures of autologous blood plasma and serum. Serum can be considered as plasma without fibrinogen, as the coagulation cascade occurs during serum extraction \cite{issaq2007serum,yu2011differences}. The concentration of fibrinogen along with other usual blood plasma parameters was measured by standard methods in the Clinical Chemistry Laboratory of Saarland University Hospital (Homburg, Germany).

\subsection{Microscopic cell structures}

To rationalize dependencies of the macroscopic sedimentation parameters on the fibrinogen concentration, experiments and numerical simulations characterizing the microscopic structure of the gel are performed. 

\subsubsection{Experiments with Quasi-2D Sedimented Structures}
\label{sec:quasi_2D_setup}

To assess structural properties of aggregated erythrocytes as a function of aggregation strength (or equivalently fibrinogen concentration), erythrocytes were allowed to settle in a pillbox-shaped microscope chamber with an inner diameter of 5 mm and a height of 1.5 mm. The initial volume fraction of $0.3\%$ was selected very low, but high enough such that after sedimentation, a quasi-2D percolating aggregate is formed at the bottom of the chamber. Note that for these experiments, phosphate-buffered saline (PBS) solutions with fibrinogen (from 0 to 550 mg/dl) were used instead of the plasma/serum mixture, in order to minimize possible donor-based variabilities. Fibrinogen used was from human plasma, provided by Sigma Aldrich (St Louis, Missouri, USA) as a powder stabilized with NaCl, and the required amount of distilled water was incorporated to the solutions to adjust osmolarities of the suspension to 290 mOsm. The final hematocrit of a $8\, \mathrm{\upmu m}$-thick monolayer of erythrocytes at the chamber bottom is approximately $\phi=0.56$, which provides a reasonable approximation for the hematocrit of the quasi-2D layer of erythrocyte gel.

\subsubsection{Numerical Simulations: Model and Methods}
\label{sec:model_methods}

Since the properties of percolating networks are known to depend on the dimensionality and the underlying particle characteristics \cite{stauffer2018introduction,manley2005gravitational,allain1995aggregation}, numerical simulations are employed to support the interpretation of the experiments and to characterize the structure of erythrocyte gels in three dimensions (3D) during sedimentation. Simulations of blood sedimentation are performed using a mesoscopic coarse-grained model of RBCs \cite{Noguchi_STV_2005,Fedosov_RBC_2010,Fedosov_SCG_2010}. Each RBC is represented by a triangulated surface with $N_\mathrm{v}=500$ vertices, whose total potential energy consists of in-plane elastic energy,
bending energy, surface-area and volume constraints as 
\begin{equation}
\label{Eq1}
    U_{\rm tot} = U_{\rm elastic} + U_{\rm bend} + U_{\rm area} + U_{\rm vol}.
\end{equation}
The elastic energy $U_{\rm elastic}$ is a sum of attractive worm-like-chain and repulsive potentials over all network edges $N_e$
\begin{equation}
    U_{\rm elastic} = \sum_{i=1}^{N_e} \left[ \frac{k_B T \ell_{\rm m}\left(3x_i^2 - 2x_i^3\right)}{4 p\left(1-x_i\right)} + \frac{k_{\rm p}}{\ell_i} \right], 
\end{equation}
where $p$ is the persistence length, $\ell_i$ is the length of edge $i$, $\ell_{\rm m}$ is the maximum edge extension, 
and $x_i = \ell_i/\ell_{\rm m}$. In the second term, $k_{\rm p}$ is the non-linear spring constant of the repulsive force coefficient, defined as a power function \cite{Fedosov_RBC_2010,Fedosov_SCG_2010}. 

The bending energy 
\begin{equation}
    U_{\rm bend} = \sum_{i=1}^{N_e} \kappa\left(1 - \cos(\theta_i-\theta_0)\right),
\end{equation}
describes the cost of membrane deformations due to curvature elasticity. Here, $\kappa$ is the bending rigidity, $\theta_i$ is the angle between two adjacent faces at the edge $i$, and $\theta_0$ represents the spontaneous curvature.  

The third and fourth terms in Eq.\,(\ref{Eq1}) correspond to constraints for surface area (both local and global) and volume, and 
are expressed as
\begin{flalign}
    U_{\rm area} &= \frac{k_{\rm a}\left(A-A_0\right)^2}{2A_0} + \sum_{i=1}^{N_{\rm f}}\frac{k_{\rm d}\left(A_i-A_i^0\right)^2}{2A_i^0},\\ \nonumber
U_{\rm vol} &= \frac{k_{\rm v}\left(V-V_0\right)^2}{2V_0},
\end{flalign}
where $k_{\rm a}$, $k_{\rm d}$ and $k_{\rm v}$ are local area, total surface area, and volume constraint coefficients, respectively. 
$N_f$ is the number of faces, $A_i$, $A$, and $V$ are the instantaneous area of face $i$, total area, and volume, respectively, and 
the quantities with a zero sub- or super-script represent their targeted values.   

\begin{figure*}[htb]
\includegraphics[width=1.0\textwidth]{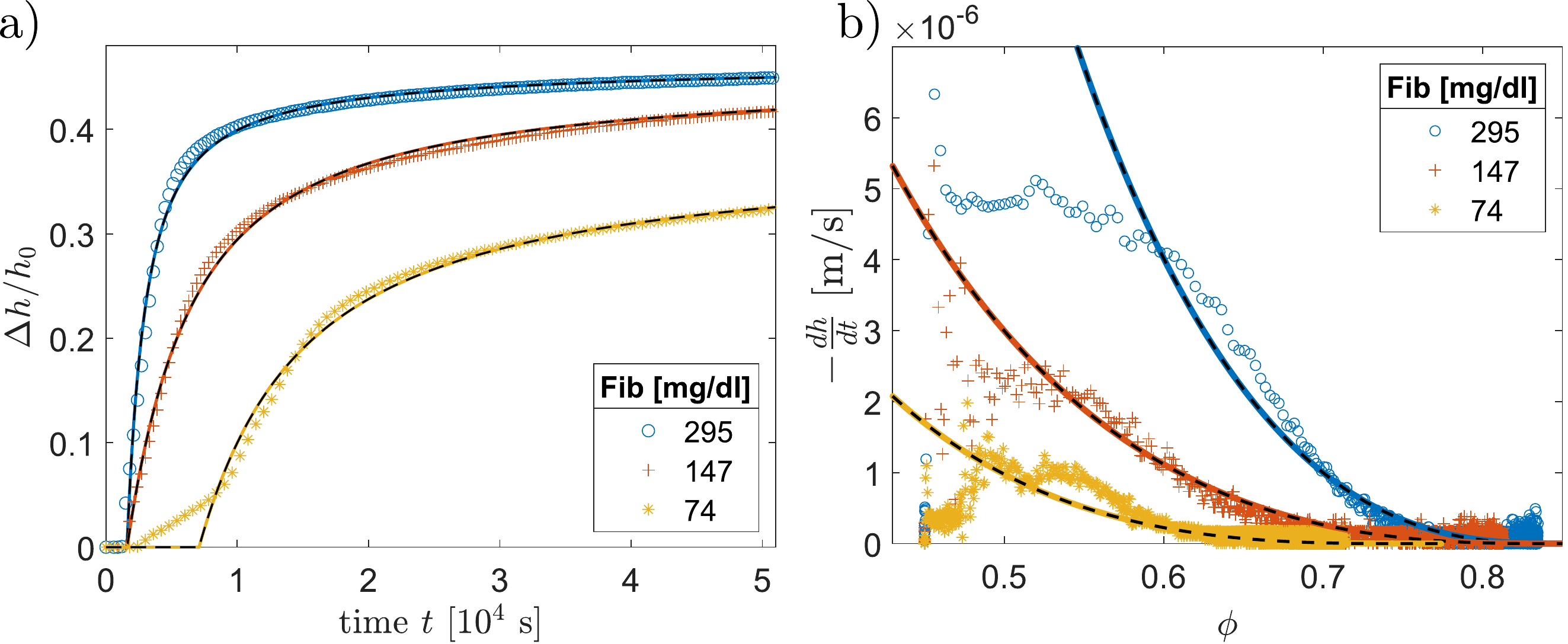}
\caption{\label{FibExpCurves} (a) Time dependence of height of the interface between dense erythrocyte suspension and cell-free plasma, for various fibrinogen concentrations. The curves show the relative variation $\Delta h/h_0=(h_0-h(t))/h_0$ which corresponds to the height of the cell-free plasma layer. The symbols are experimental data, while the curves are fits from the theoretical model for $h(t)$ in Eq.\,(\ref{FinalEq}). (b) Associated interface speed as a function of the  volume farction $\phi(t)=\phi_0 h_0/h(t)$. The symbol data are computed from a smoothing spline of the $\Delta h$ measurements, while the curves correspond to the expected speed from the fit parameters in Eq.\,(\ref{VelOne}).}
\end{figure*}

Aggregation interactions between RBC membranes is implemented via the Lennard-Jones potential 
\begin{eqnarray}
U_\mathrm{LJ} = 
\begin{cases}
4\epsilon \left[ \left(\frac{\sigma}{r}\right)^{12} -  \left(\frac{\sigma}{r}\right)^6\right]\,\,\,\textrm{for}\,\,\,r < r_\textrm{c}\\
0\,\,\,\textrm{for}\,\,\,r \geq r_\textrm{c},
\end{cases}
\end{eqnarray}
where $\epsilon$ is the aggregation strength and $\sigma=0.6~\mu$m characterizes the excluded-volume distance between RBCs. Here, the distance $\sigma$ is essentially determined by the resolution of the membrane discretization, i.e. the bond length of the membrane triangulation.
The cut-off distance $r_\textrm{c}$ is selected to be $1.2~\mu$m. To calibrate the aggregation strength  $\epsilon$, several simulations are performed, in which two RBCs are first placed close to each other and let to aggregate. Then, these RBCs are pulled apart with a force applied in the normal direction until they detach from each other, so that the force required for their detachment is determined \cite{Fedosov_PBV_2011}. For aggregation forces between 
two RBCs in the range of $2-6~\textrm{pN}$, $\epsilon$ value is in the range of $2-5~\kbt$ (e.g., $F_{\rm detach}\simeq 3.6~\textrm{pN}$
for $\epsilon=2.5~\kbt$ and $F_{\rm detach} \simeq 6~\textrm{pN}$ for $\epsilon=4.4~\kbt$). Such aggregation forces are similar to those measured experimentally in autologous blood plasma with optical tweezers \cite{ermolinskiy2020effect}. In such conditions, the total aggregation energy between two cells in the simulations is then of the order of $F_{\rm detach}\sigma\sim 6.\,10^2-9.\,10^2\,\kbt$.

In simulations, a fixed number of RBCs (depending on hematocrit) is distributed in a fluid within a simulation domain of $(50~\microm)^3$ with periodic boundary conditions in all directions. The fluid is modeled by the smoothed dissipative particle dynamics (SDPD) method, which is derived through a Lagrangian discretization of the Navier-Stokes equations \cite{Espanol_SDPD_2003,Mueller_SDPD_2015}. RBC properties correspond to average characteristics of a healthy RBC with a membrane area $A_0 = 133~\mu\textrm{m}^2$, cell volume $V_0=93~\mu\textrm{m}^3$, shear modulus $\mu=4.8~\mu\textrm{N/m}$, and bending rigidity $\kappa=70~\kbt=3\times 10^{-19}~\textrm{J}$ \cite{Evans_MTB_1980,Evans_BEM_1983,Dao_RBC_2003}.
This leads to a RBC reduced volume of $V^*=6V_0/\left( \pi D_r^3 \right) \approx 0.64$, where $D_r=\sqrt{A_0/\pi} = 6.5~\mu\textrm{m}$ is the effective RBC diameter. Note that the stress-free shape of a RBC elastic network is assumed to be the biconcave shape of a RBC with $V^* = 0.64$.
In addition to aggregating RBC suspensions, several cases without aggregation interactions are also considered by setting $r_\textrm{c}=2^{1/6}\sigma$ such that only repulsive forces between RBC vertices are present.

To mimic cell sedimentation, a constant force $F_\textrm{RBC}$ is applied to all membrane vertices of RBCs along the negative $z$ direction and force $F_\textrm{fl}$ in the opposite direction is applied to all fluid particles. Since the numbers of RBC vertices and fluid particles differ, the forces applied on each type of particles are also different. However, the total force in the simulation domain remains zero. In this way, the flow resistance of porous-like RBC structures for different simulation parameters can be measured. The pressure difference over height $z$ can be calculated as $\Delta P/ z = \rho_\textrm{RBC}\,F_\textrm{RBC} - \rho_\textrm{fl}\,F_\textrm{fl}$. This pressure difference results in a net sedimentation velocity $v = v_\textrm{RBC} - v_\textrm{fl}$. 
It is important to note that we are considering here the case of small sedimentation velocities, so that the shape of individual RBCs is only weakly affected by the flow forces \cite{peltomaeki2013}. An illustrative movie from the simulations can be found as Supplemental Movie S1 \cite{SuppMat}.

The calculation of characteristic pore sizes which describe the permeability of suspended RBC structures  is performed as follows. For each simulation trajectory, we take multiple $xy$ slices (at constant $z$) over several time intervals. For each slice, multiple straight lines are drawn in the $x$ and $y$ directions. The lengths of these lines outside the RBCs determine the lengths of void spaces and their average corresponds to an average pore size (see Supplemental Figure 1 for illustration \cite{SuppMat}). 

\section{Results}
\label{Results}

\subsection{Macroscopic Sedimentation Measurements}
\label{RawRes}

Figure \ref{FibExpCurves}(a) shows the time dependence of the erythrocyte column height during the sedimentation process for various levels of fibrinogen. The sedimentation of erythrocytes becomes slower when the concentration of fibrinogen decreases. The lines in Fig.~\ref{FibExpCurves}(a) represent the corresponding fits using the theoretical model from Eq.\,(\ref{FinalEq}).(Corresponding fit parameters, for the respective fibrinogen concentrations of $\left\{74,147,295\right\}\,\mathrm{mg/dl}$ are $\gamma=\left\{0.59,0.69,0.21\right\}$, $\phi_m=\left\{0.75,0.845,0.85\right\}$ and $t_0=\left\{7.1,1.6,1.6\right\}.10^3\,\mathrm{s}$, respectively.) The theoretical curves generally show a good agreement with experimental data, except at short times when the sedimentation starts. The beginning of sedimentation is not well captured by our model, because gels may require some structural rearrangements (e.g., dynamic formation of fluid channels) before macroscopic sedimentation can be observed \cite{derec2003rapid,allain1995aggregation,gopalakrishnan2006linking}. For instance, the curve with fibrinogen concentration $74\,\mathrm{mg/dl}$ exhibits first a slow sedimentation velocity, which then suddenly increases. Presumably, a slow settling of the gel occurs before larger fluid channels appear and significantly accelerate sedimentation in some cases \cite{derec2003rapid}. In this respect, the delay time $t_0$ characterizes the time required for 
cells to re-organize and establish dynamic fluid channels which enable fast sedimentation \cite{teece2014gels,buscall2009towards,gopalakrishnan2006linking,padmanabhan2018gravitational,bartlett2012sudden}.

\begin{figure*}[htb]
\includegraphics[width=0.8\textwidth]{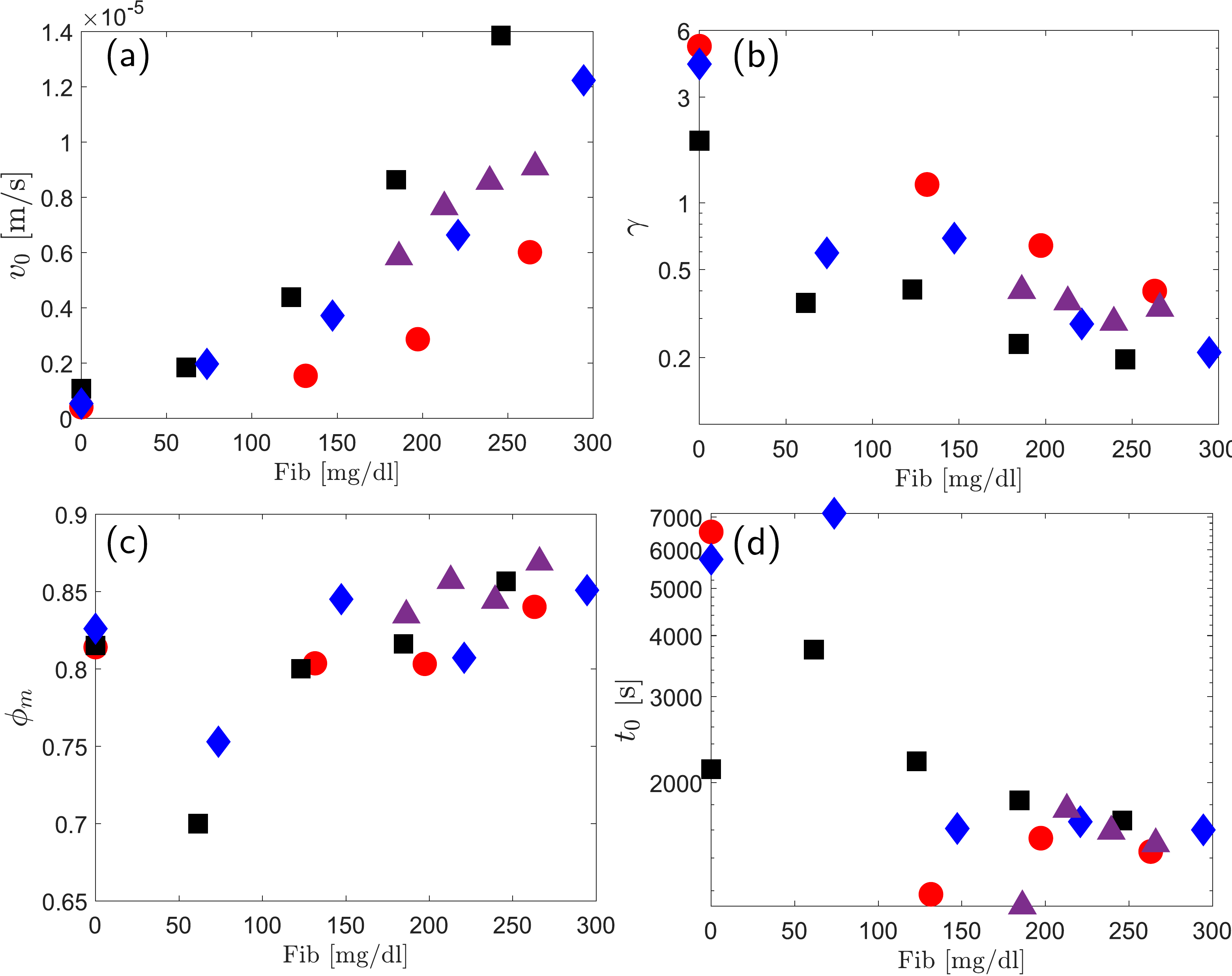}
\caption{\label{FibParamCurves} Dependence of the model parameters on the fibrinogen concentration for different blood donors. Different symbols and colors indicate different donors. (a) Initial sedimentation velocity at $t_0$ computed using Eq.\,(\ref{VelOne}) with $\phi=\phi_0=0.45$. (b) Dimensionless characteristic time $\gamma$ of the sedimentation process. (c) The final volume fraction $\phi_m$. (d) The delay time $t_0$ before the sedimentation starts.}
\end{figure*}

Figure \ref{FibExpCurves}(b) presents sedimentation velocities as a function of calculated $\phi(t)$ from height measurements, for different fibrinogen concentrations. The speed of the interface between dense erythrocyte suspension and the cell free plasma extracted from experimental measurements (symbols) show strong fluctuations, especially at the beginning of the sedimentation process or when $\phi \sim \phi_0$. However, we slightly smoothed the experimental data to calculate the numerical derivative. As result we obtain speeds that follow well the average trend given by the theoretical curves of $dh/dt$ from Eq.\,(\ref{VelOne}).    

As mentioned before, the Eq.\,(\ref{FinalEq}) allows the extraction of the model parameters from sedimentation experiments, which are shown in Fig.~\ref{FibParamCurves} along the initial speed of the interface between packed erythrocytes and the cell-free plasma. Data from various donors are shown by different symbols and colors. The initial sedimentation speed $v_0$ right at the beginning of the sedimentation process (i.e. at $\phi=\phi_0=0.45$) increases with an increase in the fibrinogen concentration [see Fig.~\ref{FibParamCurves}(a)]. Here, Eq.\,(\ref{VelOne}) is employed to compute $v_0$ instead of raw data because the relative variation in the instant velocity is of the order of magnitude of the slowest velocities observed. Even though there is some variation in $v_0$ for different donors, the common trend of increasing $v_0$ with fibrinogen concentration remains consistent. These donor-related differences are likely due to differences in the concentration of other blood proteins beside fibrinogen, since they also influence RBC aggregation and sedimentation properties, but in a smaller extent.

Consistently with the velocity increase, the dimensionless characteristic time $\gamma$ of sedimentation in Fig.~\ref{FibParamCurves}(b) decreases overall with increasing fibrinogen concentration. This can be rationalized by the fact that characteristic channel sizes (i.e., the parameter $R$ in the model) within erythrocyte structures become larger with increasing fibrinogen concentration, as will be shown later.

\begin{figure*}[ht!b]
\includegraphics[width=1.0\textwidth]{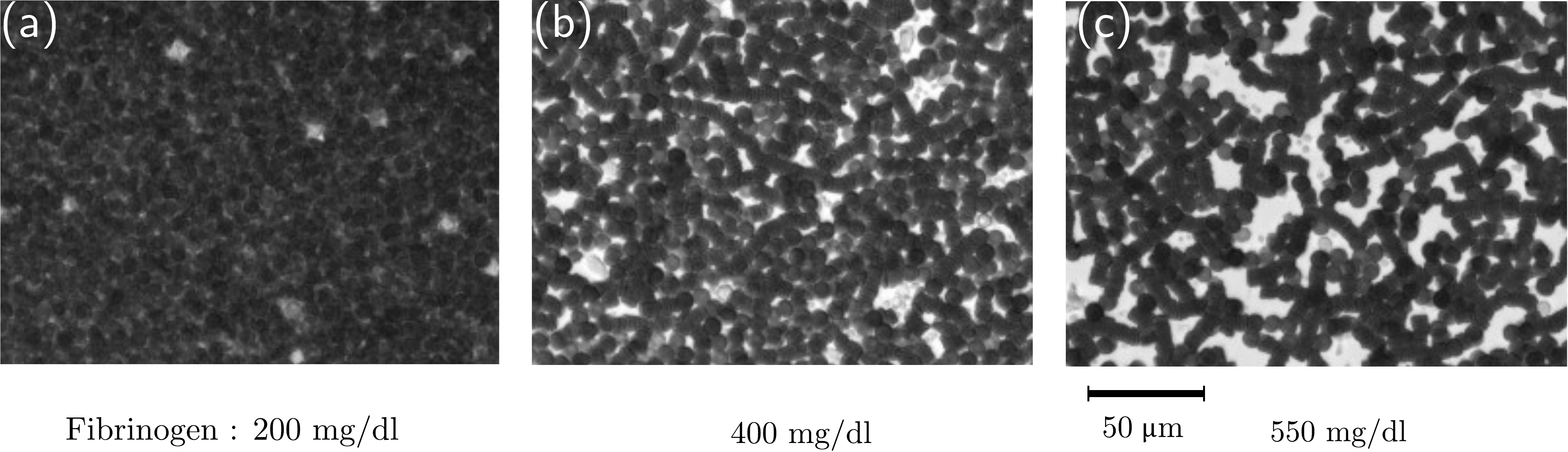}
\caption{\label{FigHoles} Pictures of 2D cell networks obtained for various fibrinogen concentrations in PBS, after sedimentation to the bottom of a microscopy chamber. The final hematocrit at the bottom of the chamber within a $8\,\mathrm{\upmu m}$ height monolayer of erythrocytes is equal to approximately $\phi=0.56$. Qualitative differences in aggregate geometries and characteristic sizes of the pores are clearly visible for different fibrinogen concentrations. Although the volume fraction is the same for each measurement, the total 2D pore area increases from (a) to (c) because more cells are seen from the side when incorporated in a rouleau, while most of the cells lie flat at the bottom of the container in panel (a).}
\end{figure*}

Figure~\ref{FibParamCurves}(c) shows that the final volume fraction $\phi_m$ of sedimented erythrocytes is nearly independent of fibrinogen concentration and mostly lies within the range of $0.84\pm0.05$. A very slight increase in $\phi_m$ with the fibrinogen concentration implies that an increasing attraction strength generates a slightly better compaction of flexible cells. However, this trend is within the order of the variations found for various donors, and further measurements would be required to rigorously support such a trend. In any case, changes in $\phi_m$ cannot explain the observed variations in the sedimentation velocity for different donors.

The delay time $t_0$ in Fig.~\ref{FibParamCurves}(d) decreases with increasing fibrinogen concentration, which indicates that stronger attractive interactions accelerate the rearrangement of initial erythrocyte structures that leads to the establishment of fluid channels. Interestingly, the dependence of $t_0$ on particle interactions is exactly opposite for suspensions of isotropic rigid colloids, such that a strong attraction stabilizes such gels \cite{teece2014gels,gopalakrishnan2006linking}. In these studies, the main proposition is that $t_0$ characterizes the time required to break existing 'bonds' within an initially stable gel. This seems to be different for a suspension of RBCs, where we see that attractive interactions can actually accelerate structural changes within the RBC gel. Here, we can partially exclude the diffusion-limited process of initial structure formation, as it is expected to occur on a significantly shorter time scale than the delay time $t_0$. For example, the characteristic time for doublet formation by diffusion-limited aggregation can be estimated as $t_D=\pi r^3_{\mathrm{RBC}}\eta/(\phi\kbt)\sim 10^2~\mathrm{s}$ \cite{russel1989colloidal}, which is an order of magnitude smaller than $t_0$ values in Fig.~\ref{FibParamCurves}(d). Furthermore, the initial diffusion-limited aggregation should not depend significantly on the aggregation strength between particles. Thus, we hypothesize that aggregation interactions can destabilize initial gel structure formed by flexible RBCs, which leads to structure rearrangement and the apparition of fluid-filled cracks. Interestingly, $t_0$ seems to have little or no effect on the the sedimentation velocity. Note that our $t_0$ measurements saturate at around 1500 s for fibrinogen concentrations larger than about 100 mg/dl, which is likely the minimal time required for the rearrangement of initial cell network. 

\subsection{Microscopic Cellular Structures}
\label{MicroExp}

\subsubsection{2D Structures of Sedimented RBCs}

Sedimentation speed should directly correlate with the permeability of erythrocyte gel, which can be characterized by the size of voids (or pores) between the cells. In experiments, it is difficult to image such voids in 3D within an ESR tube filled with blood at high hematocrit. Therefore, to assess the dependence of pore sizes within aggregated erythrocyte structures on the attraction strength, we turned to a reduced quasi-2D experiment, in which a small amount of erythrocytes is allowed to settle to the bottom of a pillbox-shaped microscope chamber, as described in Sec.~\ref{sec:quasi_2D_setup}. Typical resulting gel conformations are presented in Fig.~\ref{FigHoles}. An increase in sedimented gel porosity is also clearly visible, which shows that erythrocytes form long rouleaux structures in strongly aggregating media (i.e., at high fibrinogen concentrations).

\begin{figure}[ht!b]
\includegraphics[width=0.5\textwidth]{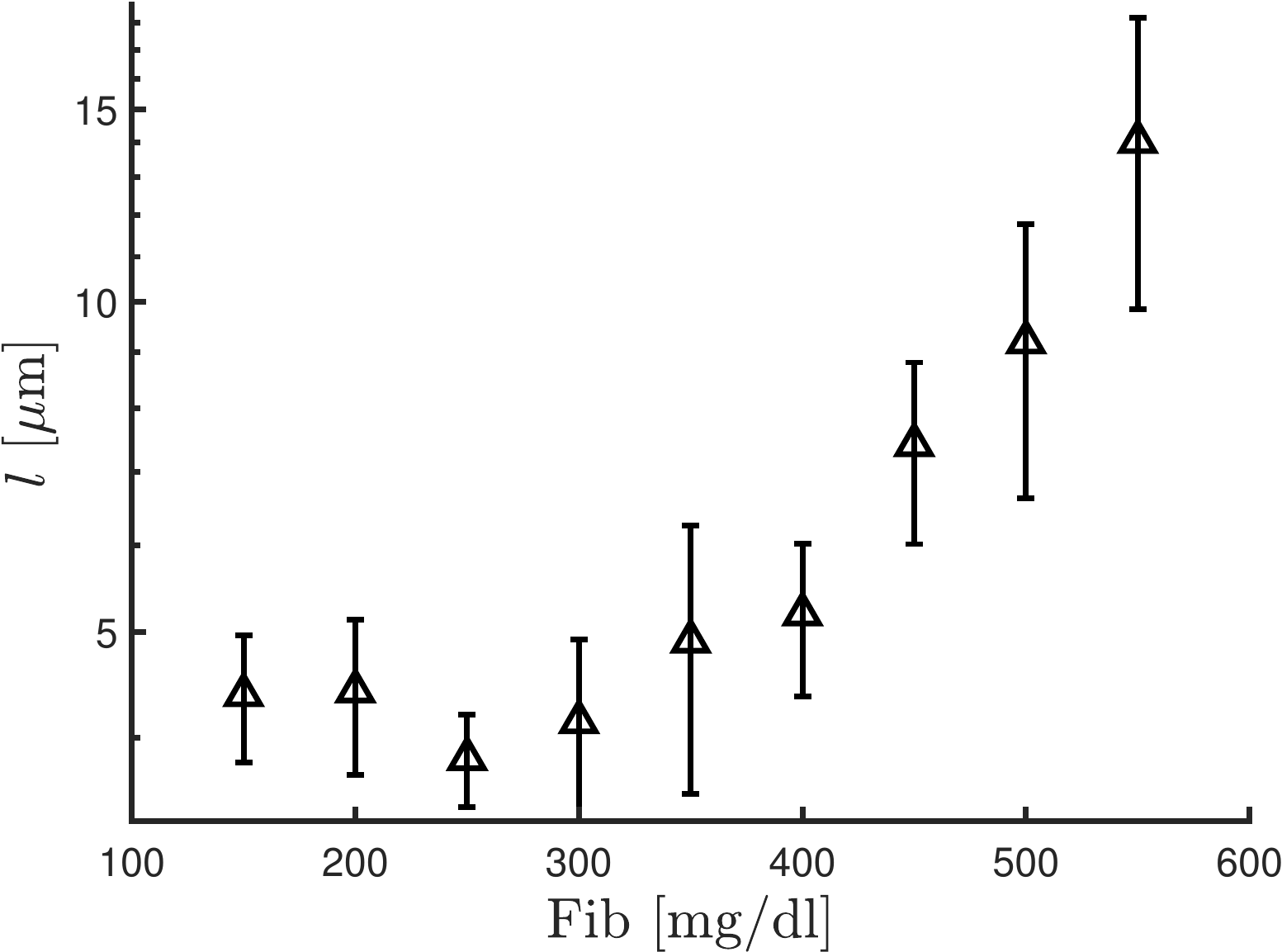}
\caption{\label{GraphHoles} Mean pore sizes within a quasi-$2D$ percolating network of erythrocytes at the bottom of the observation chamber as a function of fibrinogen concentration in PBS. A trend of increasing pore size with increasing fibrinogen concentration is observed at fibrinogen concentrations above 300 mg/dl. The bars represent standard deviations of the average values measured in various samples from different donors.}
\end{figure}

Figure \ref{GraphHoles} shows average pore sizes, defined as the square root of the pores area, as a function of fibrinogen concentration. An increase in the pore size with increasing attraction strength is in qualitative agreement with the hypothesis that the parameter $R$ should increase with the fibrinogen concentration. Although the volume fraction is the same for each measurement, the total 2D pore area then increases because more cells are seen from the side when incorporated in a rouleau, while most of the cells lie flat at the bottom of the container in Fig.\,\ref{GraphHoles}(a).  Note that the data for pore sizes in Fig.~\ref{GraphHoles} exhibit a substantial deviation for various donors, summarized through the error bars showing standard deviation for the measured average size of various samples. Such donor-dependency indicates that erythrocyte properties might also affect the magnitude of void sizes. This is not unexpected, as for colloidal suspensions, the properties of percolating gel network are known to strongly depend on the underlying geometrical and dimensional properties of suspended particles \cite{stauffer2018introduction,manley2005gravitational,allain1995aggregation}. 

\begin{figure*}[ht!b]
\includegraphics[width=0.8\textwidth]{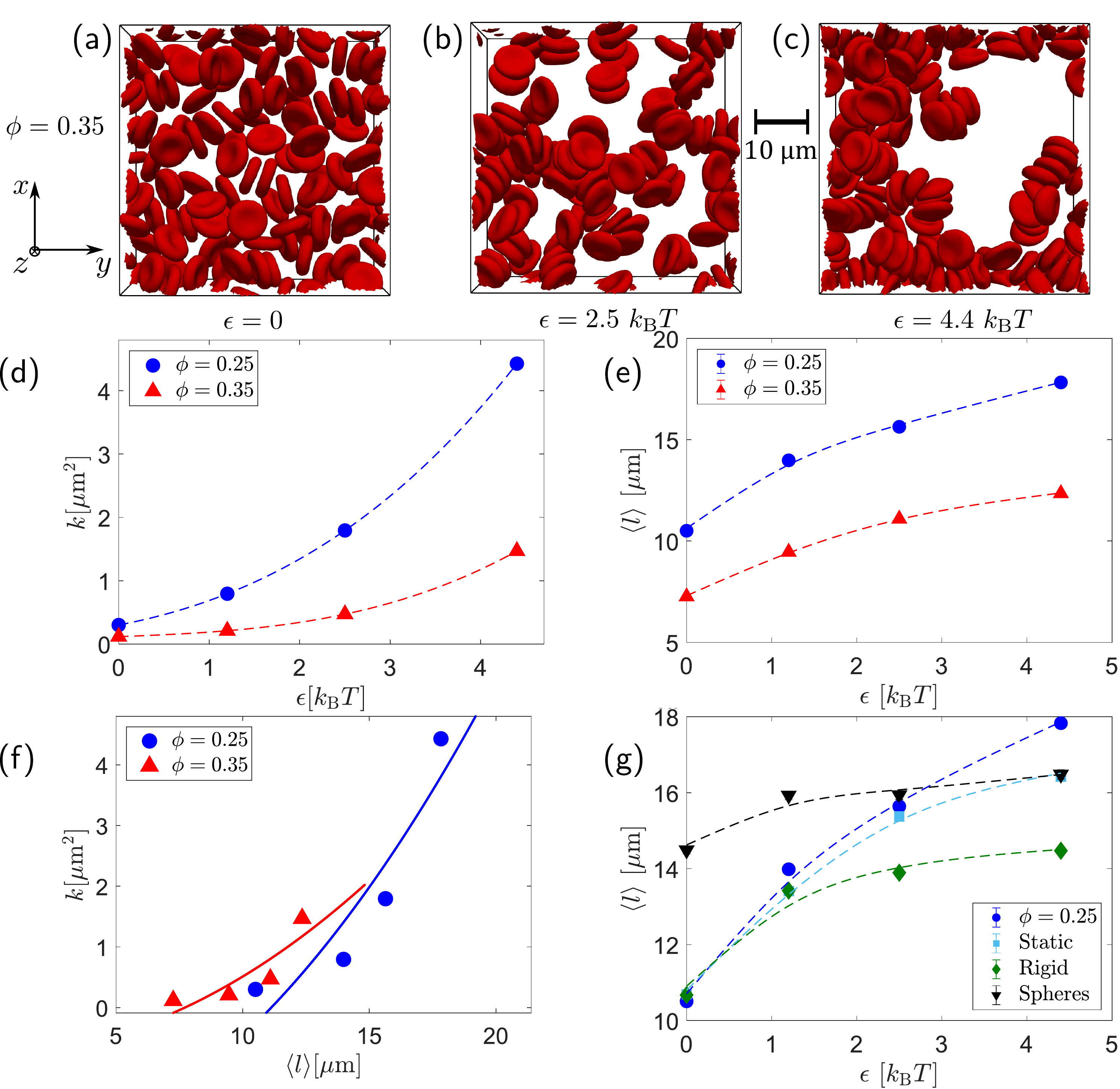}
\caption{\label{GraphsEnergy} Effect of interaction strength on the structure and permeability of erythrocyte aggregates from simulations. (a)-(c) Gelation process from hydrodynamic simulations for different interaction strengths $\epsilon=\left\{0,2.5,4.4\right\}\,\kbt$. All snapshots are for volume fraction $\phi=0.35$, and show a $10 \microm$-thick layer in $z$ with a cross-section of $50\microm\times50\microm$ in $x-y$. (d) Permeability coefficient, computed as $k = (1-\phi)v\eta/\frac{\partial P}{\partial z}$, as a function of the interaction strength $\epsilon$. The data are displayed for two different hematocrits, and the lines are a guide to the eye. (e) The average pore size $\langle l\rangle$ from simulations for varying interaction strength $\epsilon$. Errors on the average are typically smaller than the symbol sizes. The corresponding distributions of pore sizes are shown in Supplemental Figure 2 \cite{SuppMat}. The lines are a guide to the eye. (f) Permeability coefficient $k$ as a function of the mean pore size $l$ obtained by varying the interaction strength. The lines are quadratic fits ($k=A+B \langle l\rangle^2$ with $A$ and $B$ being fit parameters), which follow the scaling in the Carman-Kozeny relationship ($k\propto a^2$). (g) The average pore size $\langle l\rangle$ for different suspension conditions, including a static (no flow) case with deformable RBCs and a suspension of rigid RBCs and spherical particles. All curves are for a volume fraction of $\phi=0.25$. The lines are a guide to the eye. Errors on the average are smaller than the symbol sizes.}
\end{figure*}

\begin{figure*}[!htbp]
    \includegraphics[width=0.9\linewidth]{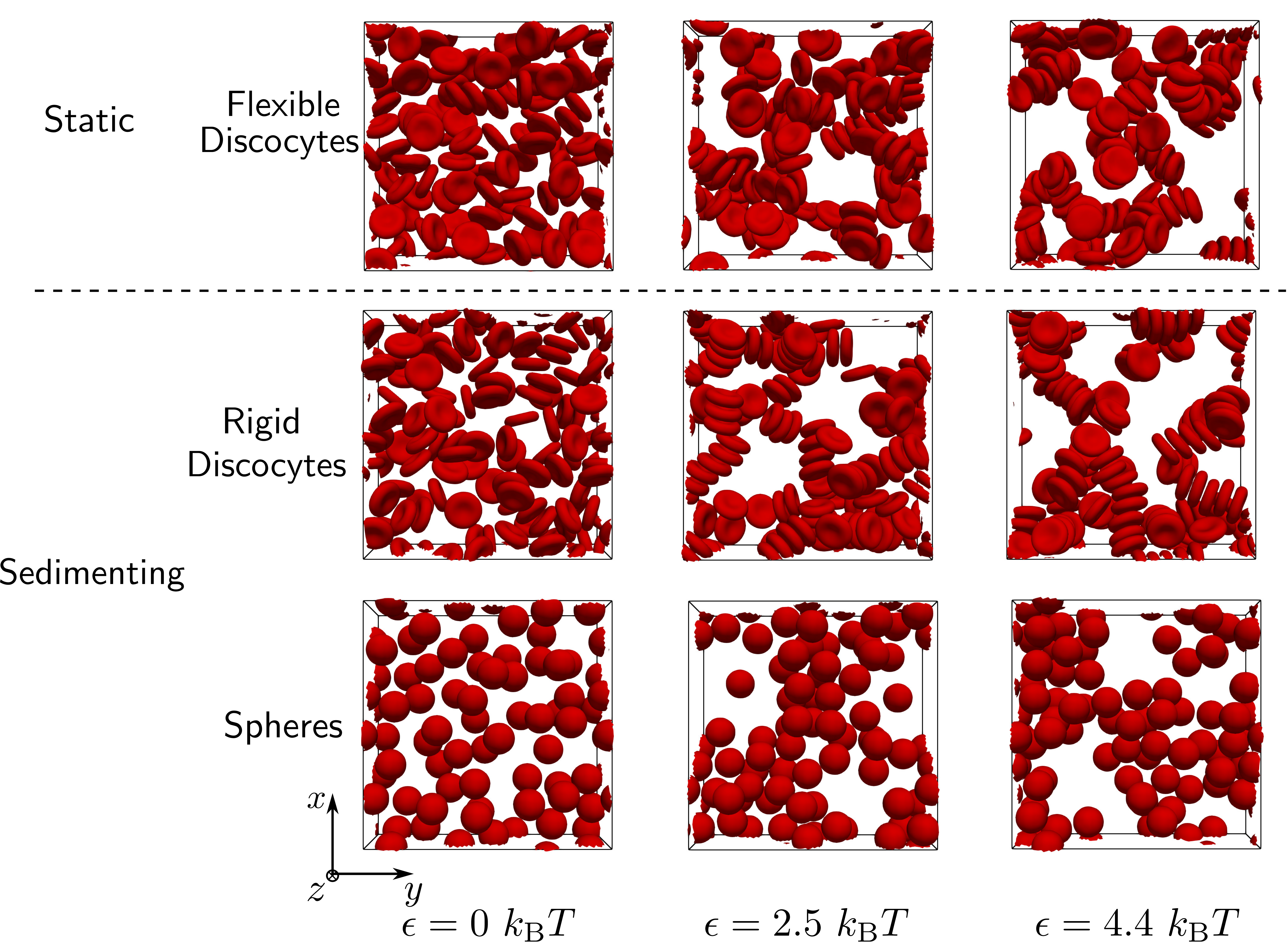}
    \caption{\textbf{Structures obtained in various conditions.} All snapshots are $15 \microm$ thick along $z$, and $50\microm\times50\microm$ in $x-y$. All snapshots are taken at volume fraction $\phi=0.25$. Each snapshot is a slice of aggregates obtained for the interaction energies $\epsilon=0\,\kbt$, $2.5\,\kbt$ and $4.4\,\kbt$, respectively. As one can see, the main difference between spheres and discocytes is the tendency of discocytes to form rouleaux. This local geometry enhances the branching of the particles and allows for an earlier percolation. The visible gaps between discocytes within a given rouleau in the pictures come from the excluded-volume distance $\sigma$ used in the simulation to ensure its stability.
}
    \label{SuppFig4}
\end{figure*}

\subsubsection{Simulations of Erythrocyte Sedimentation}
\label{MicroSimu}

\begin{figure}[ht!b]
\includegraphics[width=0.4\textwidth]{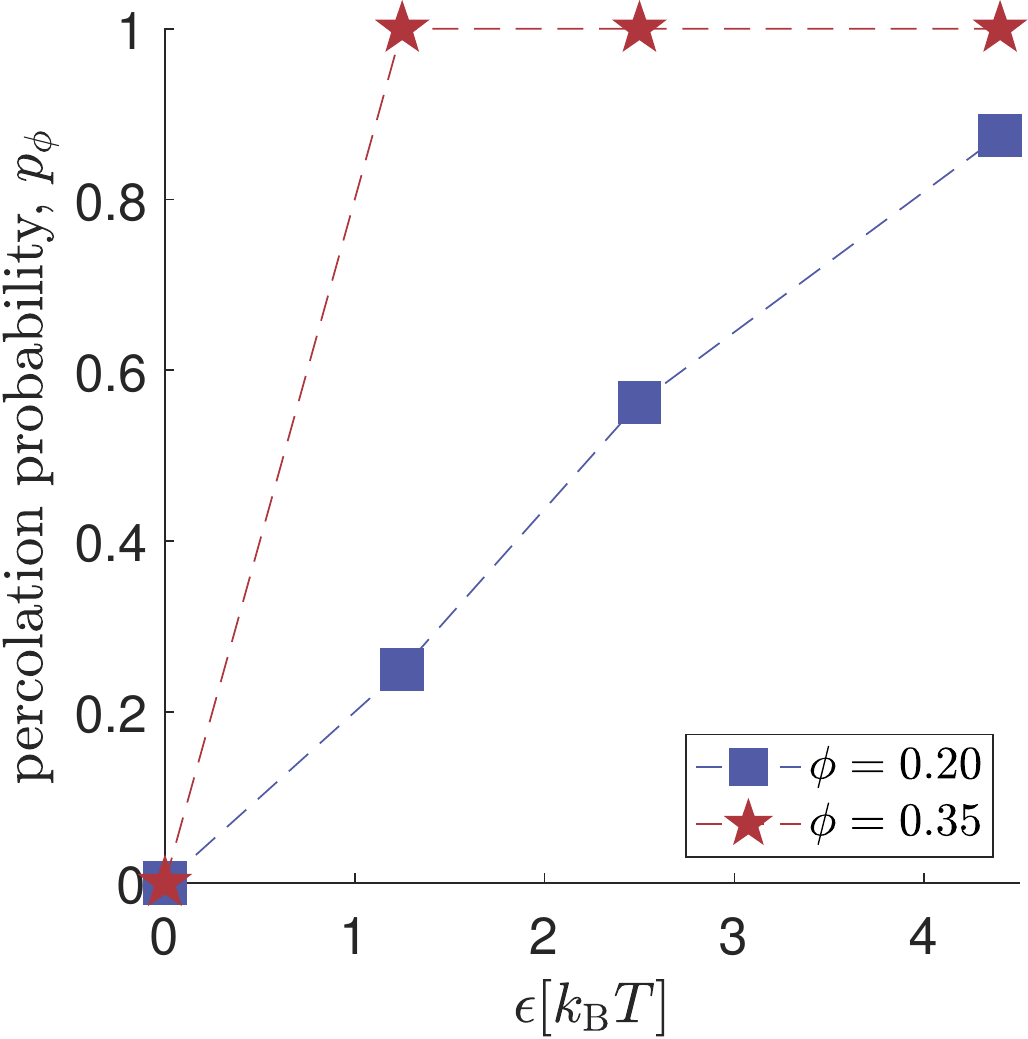}
\caption{\label{PeroclProb} Percolation probability of erythrocyte aggregates in the sedimentation simulations. In case of no aggregation, the cells sediment separately and no percolation is observed. For $\phi=0.35$, percolation already takes place at $\epsilon=1.25\kbt$. For the low volume fraction of $\phi=0.2$, the percolation probability increases gradually with the interaction strength.}
\end{figure}

To understand the relation between micro-structural properties of a RBC suspension and its sedimentation, flow simulations are performed for various hematocrits and aggregation interactions, as described in Sec.~\ref{sec:model_methods}. Characteristic simulation snapshots for three attraction strengths $\epsilon$, corresponding to three different fibrinogen concentrations, are shown in Fig.~\ref{GraphsEnergy}(a)-(c). In order to obtain a reasonable computation time, we used volume fractions of $0.25$ and $0.35$, which are smaller than characteristic values in experiments, but still produce percolating aggregates \cite{JointLetter}. The snapshots clearly demonstrate that an increase in $\epsilon$ leads to stronger clustering of RBCs and larger aggregates. Geometrical analysis of these structures (see Section~\ref{sec:model_methods} for details) allows the extraction of the permeability coefficient $k$ and the average pore size $\langle l \rangle$. Figure~\ref{GraphsEnergy}(d) presents the permeability coefficient which increases with increasing interaction strength $\epsilon$, at least in the range of physiological attractive interactions. The characteristic pore size $\langle l\rangle$ in Fig.~\ref{GraphsEnergy}(e) also consistently increases with the interaction strength. 

An increase in the permeability of the erythrocyte gel with increasing pore size is explicitly shown in Fig.~\ref{GraphsEnergy}(f). More importantly, the dependence of $k$ on $\langle l \rangle$ is nearly quadratic (i.e., $k\propto \langle l \rangle^2$), which is consistent with the the Carman-Kozeny relationship used in the theoretical model (see Sec.~\ref{ThRecal} and the joint letter \cite{JointLetter}). In agreement with experimental observations, the increase in the pore size is related to the formation of long network-like rouleaux aggregates (see Fig.~\ref{FigHoles}). This argument is also consistent with distributions of $l$ (see Supplemental Figure 2 \cite{SuppMat}) which show that an increasing interaction strength mainly leads to an increase in the number of large pores within the gel network. The combination of these observations explain how an increase in fibrinogen concentration during inflammation results in a faster collapse (or sedimentation) of the erythrocyte gel. Indeed, they show that an increase of erythrocytes attraction leads to a higher permeability through geometrical effects, which are then responsible for the higher sedimentation rate. This then reconciles the usual diagnostic explanation with the gel-collapse model.

In addition to the effect of interaction strength on the size of pores within erythrocyte gel, it is interesting to look at the importance of fluid stresses, cell elasticity and cell shape for sedimentation. Figure~\ref{GraphsEnergy}(g) presents average pore sizes for the cases with no fluid flow, and for rigidified RBCs and spherical particles. The snapshots of corresponding structures are shown in Fig. \ref{SuppFig4}. The comparison of the $\phi=0.25$ curve (with sedimentation flow) with the "static" curve in Fig.~\ref{GraphsEnergy}(g) shows that $\langle l \rangle$ is only slightly affected by fluid stresses. This is likely due to the fact that fluid flow during sedimentation is generally slow, and thus, the corresponding fluid stresses do not induce significant rearrangements within aggregated network structures. Interestingly, the largest difference in $\langle l \rangle$ occurs at the largest $\epsilon$, which suggests that structures in highly aggregating media might be more fragile. Note that one of the limitations of our simulations is a relatively small simulation domain, since aggregated structures may have differences in stability over different length scales. 

We also consider for comparison the case of rigidified cells, whose bending and shear moduli of the membrane are set to be ten times larger than those of normal RBCs. The results in Fig.~\ref{GraphsEnergy}(g) demonstrate that even though $\langle l \rangle$ still increases with $\epsilon$ for rigidified RBCs, the slope of the curve nearly vanishes at large enough attraction strength. This means that for $\epsilon \gtrsim \kbt$, membrane deformability plays a dominant role for the pore size and the speed of sedimentation. For instance, this is likely the main reason why the ESR is very low in acanthocytosis disease, where an increased rigidity and shape changes in RBCs are frequently observed \cite{Darras2020Erythrocytes}. For comparison, Fig.~\ref{GraphsEnergy}(g) also presents pore sizes for a suspension of sedimenting spheres. At low $\epsilon$, $\langle l \rangle$ for the suspension of spheres is larger than that for RBCs, leading to a faster sedimentation. As $\epsilon$ increases, the average pore size quickly saturates and becomes independent of $\epsilon$, which is consistent with the effect of particle rigidity on sedimentation discussed above. Furthermore, this observation is in agreement with properties of rigid-colloid suspensions, such that spherical particles form fractal aggregates whose fractal dimensions depend on the aggregation regime, but not on the magnitude of the attractive force \cite{gelb2019permeability}. As a result, the dependence of ESR on fibrinogen concentration for erythrocyte suspensions is governed by the combination of anisotropic shape and membrane flexibility of these cells. 

Percolation of erythrocyte aggregate networks already takes place at a hematocrit of about $\phi=0.2$ for normal aggregation levels of 
$\epsilon = 2-3\,\kbt$ \cite{JointLetter}. Figure \ref{PeroclProb} demonstrates that the interaction strength modifies the percolation probability $p_{\phi}$ at low volume fractions. The percolation probability $p_{\phi}$ is defined as the fraction of time when the largest cell cluster within the computational domain percolates through both $x$ and $y$ directions (excluding the sedimentation $z$ direction). For $\phi=0.2$, $p_{\phi}$ gradually increases with the interaction strength. This dependence of $p_{\phi}$ is also consistent with the observation that the pore size increases with increasing $\epsilon$, as network-like aggregates grow and yield larger void spaces.

\section{Summary and Conclusions}
\label{Discussion}

We have shown by a combination of experiments, theory, and simulations that an increase in erythrocyte sedimentation velocity due to an increase in aggregation interactions between eryhtrocytes can be explained through the formation of a gel structure, its permeability characteristics, and collapse dynamics. Suspensions of spherical and rigid particles show much less sensitivity of the sedimentation speed to the attraction strength, demonstrating that RBC anisotropy and flexibility govern the structure of RBC aggregates and their sedimentation behavior. These results are consistent with the erythrocyte gel hypothesis and the theoretical model of sedimentation we introduced \cite{JointLetter}. 

Our measurements also demonstrate an unexpected dependence of the delay time $t_0$ before sedimentation starts, on the aggregation strength. In particular, $t_0$ decreases with increasing $\epsilon$, so that the sedimentation process starts faster. This behavior is exactly opposite for suspensions of rigid colloids, where attractive interactions stabilise the gel and delay its sedimentation. We hypothesize that RBC membrane flexibility determines $t_0$ and reduces erythrocyte gel stability at large interaction strengths. Nevertheless, these differences in the sedimentation of suspensions of soft and rigid particles require further investigation.

In conclusion, our results show that erythrocyte suspensions should be considered as soft-colloid gels, in contrast to the common proposition of the sedimentation as large, separate cell aggregates. 
Thus, the structure and dynamics of erythrocyte suspensions can be analyzed and characterized by same physical tools and as many other colloidal suspensions with a high volume fraction. 
This observation opens up new ways to develop novel diagnostic applications in diseases (e.g., acanthocytosis) involving some modification of erythrocyte properties, as suggested in a previous study \cite{Darras2020Erythrocytes}. Importantly, RBC anisotropy and flexibility place these suspensions into a class of soft-colloid systems whose sedimentation behavior and comprehensive theoretical description are still open physical questions which have to be addressed in future research.

\bibliography{apssamp}

\providecommand{\noopsort}[1]{}\providecommand{\singleletter}[1]{#1}%
\begin{thebibliography}{56}%
\makeatletter
\providecommand \@ifxundefined [1]{%
 \@ifx{#1\undefined}
}%
\providecommand \@ifnum [1]{%
 \ifnum #1\expandafter \@firstoftwo
 \else \expandafter \@secondoftwo
 \fi
}%
\providecommand \@ifx [1]{%
 \ifx #1\expandafter \@firstoftwo
 \else \expandafter \@secondoftwo
 \fi
}%
\providecommand \natexlab [1]{#1}%
\providecommand \enquote  [1]{``#1''}%
\providecommand \bibnamefont  [1]{#1}%
\providecommand \bibfnamefont [1]{#1}%
\providecommand \citenamefont [1]{#1}%
\providecommand \href@noop [0]{\@secondoftwo}%
\providecommand \href [0]{\begingroup \@sanitize@url \@href}%
\providecommand \@href[1]{\@@startlink{#1}\@@href}%
\providecommand \@@href[1]{\endgroup#1\@@endlink}%
\providecommand \@sanitize@url [0]{\catcode `\\12\catcode `\$12\catcode
  `\&12\catcode `\#12\catcode `\^12\catcode `\_12\catcode `\%12\relax}%
\providecommand \@@startlink[1]{}%
\providecommand \@@endlink[0]{}%
\providecommand \url  [0]{\begingroup\@sanitize@url \@url }%
\providecommand \@url [1]{\endgroup\@href {#1}{\urlprefix }}%
\providecommand \urlprefix  [0]{URL }%
\providecommand \Eprint [0]{\href }%
\providecommand \doibase [0]{https://doi.org/}%
\providecommand \selectlanguage [0]{\@gobble}%
\providecommand \bibinfo  [0]{\@secondoftwo}%
\providecommand \bibfield  [0]{\@secondoftwo}%
\providecommand \translation [1]{[#1]}%
\providecommand \BibitemOpen [0]{}%
\providecommand \bibitemStop [0]{}%
\providecommand \bibitemNoStop [0]{.\EOS\space}%
\providecommand \EOS [0]{\spacefactor3000\relax}%
\providecommand \BibitemShut  [1]{\csname bibitem#1\endcsname}%
\let\auto@bib@innerbib\@empty
\bibitem [{\citenamefont {Lapi{\'c}}\ \emph
  {et~al.}(2020{\natexlab{a}})\citenamefont {Lapi{\'c}}, \citenamefont
  {Padoan}, \citenamefont {Bozzato},\ and\ \citenamefont
  {Plebani}}]{lapic2020erythrocyte}%
  \BibitemOpen
  \bibfield  {author} {\bibinfo {author} {\bibfnamefont {I.}~\bibnamefont
  {Lapi{\'c}}}, \bibinfo {author} {\bibfnamefont {A.}~\bibnamefont {Padoan}},
  \bibinfo {author} {\bibfnamefont {D.}~\bibnamefont {Bozzato}},\ and\ \bibinfo
  {author} {\bibfnamefont {M.}~\bibnamefont {Plebani}},\ }\bibfield  {title}
  {\bibinfo {title} {Erythrocyte sedimentation rate and c-reactive protein in
  acute inflammation: meta-analysis of diagnostic accuracy studies},\
  }\href@noop {} {\bibfield  {journal} {\bibinfo  {journal} {American Journal
  of Clinical Pathology}\ }\textbf {\bibinfo {volume} {153}},\ \bibinfo {pages}
  {14} (\bibinfo {year} {2020}{\natexlab{a}})}\BibitemShut {NoStop}%
\bibitem [{\citenamefont {Lapi{\'c}}\ \emph
  {et~al.}(2020{\natexlab{b}})\citenamefont {Lapi{\'c}}, \citenamefont
  {Rogi{\'c}},\ and\ \citenamefont {Plebani}}]{lapic2020Corona}%
  \BibitemOpen
  \bibfield  {author} {\bibinfo {author} {\bibfnamefont {I.}~\bibnamefont
  {Lapi{\'c}}}, \bibinfo {author} {\bibfnamefont {D.}~\bibnamefont
  {Rogi{\'c}}},\ and\ \bibinfo {author} {\bibfnamefont {M.}~\bibnamefont
  {Plebani}},\ }\bibfield  {title} {\bibinfo {title} {Erythrocyte sedimentation
  rate is associated with severe coronavirus disease 2019 (covid-19): a pooled
  analysis},\ }\href@noop {} {\bibfield  {journal} {\bibinfo  {journal}
  {Clinical Chemistry and Laboratory Medicine (CCLM)}\ }\textbf {\bibinfo
  {volume} {1}} (\bibinfo {year} {2020}{\natexlab{b}})}\BibitemShut {NoStop}%
\bibitem [{\citenamefont {McCabe}(1985)}]{mccabe1985brief}%
  \BibitemOpen
  \bibfield  {author} {\bibinfo {author} {\bibfnamefont {B.~H.}\ \bibnamefont
  {McCabe}},\ }\bibfield  {title} {\bibinfo {title} {A brief history of the
  erythrocyte sedimentation rate},\ }\href@noop {} {\bibfield  {journal}
  {\bibinfo  {journal} {Laboratory Medicine}\ }\textbf {\bibinfo {volume}
  {16}},\ \bibinfo {pages} {177} (\bibinfo {year} {1985})}\BibitemShut
  {NoStop}%
\bibitem [{\citenamefont {Taye}(2020)}]{taye2020sedimentation}%
  \BibitemOpen
  \bibfield  {author} {\bibinfo {author} {\bibfnamefont {M.~A.}\ \bibnamefont
  {Taye}},\ }\bibfield  {title} {\bibinfo {title} {Sedimentation rate of
  erythrocyte from physics prospective},\ }\href@noop {} {\bibfield  {journal}
  {\bibinfo  {journal} {The European Physical Journal E}\ }\textbf {\bibinfo
  {volume} {43}},\ \bibinfo {pages} {1} (\bibinfo {year} {2020})}\BibitemShut
  {NoStop}%
\bibitem [{\citenamefont {Baskurt}\ \emph {et~al.}(2011)\citenamefont
  {Baskurt}, \citenamefont {Neu},\ and\ \citenamefont
  {Meiselman}}]{baskurt2011red}%
  \BibitemOpen
  \bibfield  {author} {\bibinfo {author} {\bibfnamefont {O.}~\bibnamefont
  {Baskurt}}, \bibinfo {author} {\bibfnamefont {B.}~\bibnamefont {Neu}},\ and\
  \bibinfo {author} {\bibfnamefont {H.~J.}\ \bibnamefont {Meiselman}},\
  }\href@noop {} {\emph {\bibinfo {title} {Red blood cell aggregation}}}\
  (\bibinfo  {publisher} {CRC Press},\ \bibinfo {year} {2011})\BibitemShut
  {NoStop}%
\bibitem [{\citenamefont {Smallwood}\ \emph {et~al.}(1985)\citenamefont
  {Smallwood}, \citenamefont {Tindale},\ and\ \citenamefont
  {Trowbridge}}]{smallwood1985physics}%
  \BibitemOpen
  \bibfield  {author} {\bibinfo {author} {\bibfnamefont {R.}~\bibnamefont
  {Smallwood}}, \bibinfo {author} {\bibfnamefont {W.}~\bibnamefont {Tindale}},\
  and\ \bibinfo {author} {\bibfnamefont {E.}~\bibnamefont {Trowbridge}},\
  }\bibfield  {title} {\bibinfo {title} {The physics of red cell
  sedimentation},\ }\href@noop {} {\bibfield  {journal} {\bibinfo  {journal}
  {Physics in Medicine \& Biology}\ }\textbf {\bibinfo {volume} {30}},\
  \bibinfo {pages} {125} (\bibinfo {year} {1985})}\BibitemShut {NoStop}%
\bibitem [{\citenamefont {Puccini}\ \emph {et~al.}(1977)\citenamefont
  {Puccini}, \citenamefont {Stasiw},\ and\ \citenamefont
  {Cerny}}]{puccini1977erythrocyte}%
  \BibitemOpen
  \bibfield  {author} {\bibinfo {author} {\bibfnamefont {C.}~\bibnamefont
  {Puccini}}, \bibinfo {author} {\bibfnamefont {D.}~\bibnamefont {Stasiw}},\
  and\ \bibinfo {author} {\bibfnamefont {L.}~\bibnamefont {Cerny}},\ }\bibfield
   {title} {\bibinfo {title} {The erythrocyte sedimentation curve: a
  semi-empirical approach},\ }\href@noop {} {\bibfield  {journal} {\bibinfo
  {journal} {Biorheology}\ }\textbf {\bibinfo {volume} {14}},\ \bibinfo {pages}
  {43} (\bibinfo {year} {1977})}\BibitemShut {NoStop}%
\bibitem [{\citenamefont {Dorrington}\ and\ \citenamefont
  {Johnston}(1983)}]{dorrington1983erythrocyte}%
  \BibitemOpen
  \bibfield  {author} {\bibinfo {author} {\bibfnamefont {K.~L.}\ \bibnamefont
  {Dorrington}}\ and\ \bibinfo {author} {\bibfnamefont {B.~S.}\ \bibnamefont
  {Johnston}},\ }\bibfield  {title} {\bibinfo {title} {The erythrocyte
  sedimentation rate time curve: Critique of an established solution},\
  }\href@noop {} {\bibfield  {journal} {\bibinfo  {journal} {Journal of
  biomechanics}\ }\textbf {\bibinfo {volume} {16}},\ \bibinfo {pages} {99}
  (\bibinfo {year} {1983})}\BibitemShut {NoStop}%
\bibitem [{\citenamefont {Sousa}\ \emph {et~al.}(2018)\citenamefont {Sousa},
  \citenamefont {dos Santos}, \citenamefont {Magna},\ and\ \citenamefont
  {de~Oliveira}}]{sousa2018validation}%
  \BibitemOpen
  \bibfield  {author} {\bibinfo {author} {\bibfnamefont {J.~V. d.~C.}\
  \bibnamefont {Sousa}}, \bibinfo {author} {\bibfnamefont {M.~N.}\ \bibnamefont
  {dos Santos}}, \bibinfo {author} {\bibfnamefont {L.}~\bibnamefont {Magna}},\
  and\ \bibinfo {author} {\bibfnamefont {E.~C.}\ \bibnamefont {de~Oliveira}},\
  }\bibfield  {title} {\bibinfo {title} {Validation of a fractional model for
  erythrocyte sedimentation rate},\ }\href@noop {} {\bibfield  {journal}
  {\bibinfo  {journal} {Computational and Applied Mathematics}\ }\textbf
  {\bibinfo {volume} {37}},\ \bibinfo {pages} {6903} (\bibinfo {year}
  {2018})}\BibitemShut {NoStop}%
\bibitem [{\citenamefont {Rouwhorst}\ \emph {et~al.}(2020)\citenamefont
  {Rouwhorst}, \citenamefont {Schall}, \citenamefont {Ness}, \citenamefont
  {Blijdenstein},\ and\ \citenamefont {Zaccone}}]{rouwhorst2020nonequilibrium}%
  \BibitemOpen
  \bibfield  {author} {\bibinfo {author} {\bibfnamefont {J.}~\bibnamefont
  {Rouwhorst}}, \bibinfo {author} {\bibfnamefont {P.}~\bibnamefont {Schall}},
  \bibinfo {author} {\bibfnamefont {C.}~\bibnamefont {Ness}}, \bibinfo {author}
  {\bibfnamefont {T.}~\bibnamefont {Blijdenstein}},\ and\ \bibinfo {author}
  {\bibfnamefont {A.}~\bibnamefont {Zaccone}},\ }\bibfield  {title} {\bibinfo
  {title} {Nonequilibrium master kinetic equation modeling of colloidal
  gelation},\ }\href@noop {} {\bibfield  {journal} {\bibinfo  {journal}
  {Physical Review E}\ }\textbf {\bibinfo {volume} {102}},\ \bibinfo {pages}
  {022602} (\bibinfo {year} {2020})}\BibitemShut {NoStop}%
\bibitem [{\citenamefont {Guo}\ \emph {et~al.}(2011)\citenamefont {Guo},
  \citenamefont {Ramakrishnan}, \citenamefont {Harden},\ and\ \citenamefont
  {Leheny}}]{guo2011gel}%
  \BibitemOpen
  \bibfield  {author} {\bibinfo {author} {\bibfnamefont {H.}~\bibnamefont
  {Guo}}, \bibinfo {author} {\bibfnamefont {S.}~\bibnamefont {Ramakrishnan}},
  \bibinfo {author} {\bibfnamefont {J.~L.}\ \bibnamefont {Harden}},\ and\
  \bibinfo {author} {\bibfnamefont {R.~L.}\ \bibnamefont {Leheny}},\ }\bibfield
   {title} {\bibinfo {title} {Gel formation and aging in weakly attractive
  nanocolloid suspensions at intermediate concentrations},\ }\href@noop {}
  {\bibfield  {journal} {\bibinfo  {journal} {The Journal of chemical physics}\
  }\textbf {\bibinfo {volume} {135}},\ \bibinfo {pages} {154903} (\bibinfo
  {year} {2011})}\BibitemShut {NoStop}%
\bibitem [{\citenamefont {Teece}\ \emph {et~al.}(2014)\citenamefont {Teece},
  \citenamefont {Hart}, \citenamefont {Hsu}, \citenamefont {Gilligan},
  \citenamefont {Faers},\ and\ \citenamefont {Bartlett}}]{teece2014gels}%
  \BibitemOpen
  \bibfield  {author} {\bibinfo {author} {\bibfnamefont {L.~J.}\ \bibnamefont
  {Teece}}, \bibinfo {author} {\bibfnamefont {J.~M.}\ \bibnamefont {Hart}},
  \bibinfo {author} {\bibfnamefont {K.~Y.~N.}\ \bibnamefont {Hsu}}, \bibinfo
  {author} {\bibfnamefont {S.}~\bibnamefont {Gilligan}}, \bibinfo {author}
  {\bibfnamefont {M.~A.}\ \bibnamefont {Faers}},\ and\ \bibinfo {author}
  {\bibfnamefont {P.}~\bibnamefont {Bartlett}},\ }\bibfield  {title} {\bibinfo
  {title} {Gels under stress: The origins of delayed collapse},\ }\href@noop {}
  {\bibfield  {journal} {\bibinfo  {journal} {Colloids and Surfaces A:
  Physicochemical and Engineering Aspects}\ }\textbf {\bibinfo {volume}
  {458}},\ \bibinfo {pages} {126} (\bibinfo {year} {2014})}\BibitemShut
  {NoStop}%
\bibitem [{\citenamefont {Buscall}\ \emph {et~al.}(2009)\citenamefont
  {Buscall}, \citenamefont {Choudhury}, \citenamefont {Faers}, \citenamefont
  {Goodwin}, \citenamefont {Luckham},\ and\ \citenamefont
  {Partridge}}]{buscall2009towards}%
  \BibitemOpen
  \bibfield  {author} {\bibinfo {author} {\bibfnamefont {R.}~\bibnamefont
  {Buscall}}, \bibinfo {author} {\bibfnamefont {T.~H.}\ \bibnamefont
  {Choudhury}}, \bibinfo {author} {\bibfnamefont {M.~A.}\ \bibnamefont
  {Faers}}, \bibinfo {author} {\bibfnamefont {J.~W.}\ \bibnamefont {Goodwin}},
  \bibinfo {author} {\bibfnamefont {P.~A.}\ \bibnamefont {Luckham}},\ and\
  \bibinfo {author} {\bibfnamefont {S.~J.}\ \bibnamefont {Partridge}},\
  }\bibfield  {title} {\bibinfo {title} {Towards rationalising collapse times
  for the delayed sedimentation of weakly-aggregated colloidal gels},\
  }\href@noop {} {\bibfield  {journal} {\bibinfo  {journal} {Soft Matter}\
  }\textbf {\bibinfo {volume} {5}},\ \bibinfo {pages} {1345} (\bibinfo {year}
  {2009})}\BibitemShut {NoStop}%
\bibitem [{\citenamefont {Gopalakrishnan}\ \emph {et~al.}(2006)\citenamefont
  {Gopalakrishnan}, \citenamefont {Schweizer},\ and\ \citenamefont
  {Zukoski}}]{gopalakrishnan2006linking}%
  \BibitemOpen
  \bibfield  {author} {\bibinfo {author} {\bibfnamefont {V.}~\bibnamefont
  {Gopalakrishnan}}, \bibinfo {author} {\bibfnamefont {K.~S.}\ \bibnamefont
  {Schweizer}},\ and\ \bibinfo {author} {\bibfnamefont {C.}~\bibnamefont
  {Zukoski}},\ }\bibfield  {title} {\bibinfo {title} {Linking single particle
  rearrangements to delayed collapse times in transient depletion gels},\
  }\href@noop {} {\bibfield  {journal} {\bibinfo  {journal} {Journal of
  Physics: Condensed Matter}\ }\textbf {\bibinfo {volume} {18}},\ \bibinfo
  {pages} {11531} (\bibinfo {year} {2006})}\BibitemShut {NoStop}%
\bibitem [{\citenamefont {Padmanabhan}\ and\ \citenamefont
  {Zia}(2018)}]{padmanabhan2018gravitational}%
  \BibitemOpen
  \bibfield  {author} {\bibinfo {author} {\bibfnamefont {P.}~\bibnamefont
  {Padmanabhan}}\ and\ \bibinfo {author} {\bibfnamefont {R.}~\bibnamefont
  {Zia}},\ }\bibfield  {title} {\bibinfo {title} {Gravitational collapse of
  colloidal gels: non-equilibrium phase separation driven by osmotic
  pressure},\ }\href@noop {} {\bibfield  {journal} {\bibinfo  {journal} {Soft
  Matter}\ }\textbf {\bibinfo {volume} {14}},\ \bibinfo {pages} {3265}
  (\bibinfo {year} {2018})}\BibitemShut {NoStop}%
\bibitem [{\citenamefont {Bartlett}\ \emph {et~al.}(2012)\citenamefont
  {Bartlett}, \citenamefont {Teece},\ and\ \citenamefont
  {Faers}}]{bartlett2012sudden}%
  \BibitemOpen
  \bibfield  {author} {\bibinfo {author} {\bibfnamefont {P.}~\bibnamefont
  {Bartlett}}, \bibinfo {author} {\bibfnamefont {L.~J.}\ \bibnamefont
  {Teece}},\ and\ \bibinfo {author} {\bibfnamefont {M.~A.}\ \bibnamefont
  {Faers}},\ }\bibfield  {title} {\bibinfo {title} {Sudden collapse of a
  colloidal gel},\ }\href@noop {} {\bibfield  {journal} {\bibinfo  {journal}
  {Physical Review E}\ }\textbf {\bibinfo {volume} {85}},\ \bibinfo {pages}
  {021404} (\bibinfo {year} {2012})}\BibitemShut {NoStop}%
\bibitem [{\citenamefont {Manley}\ \emph {et~al.}(2005)\citenamefont {Manley},
  \citenamefont {Skotheim}, \citenamefont {Mahadevan},\ and\ \citenamefont
  {Weitz}}]{manley2005gravitational}%
  \BibitemOpen
  \bibfield  {author} {\bibinfo {author} {\bibfnamefont {S.}~\bibnamefont
  {Manley}}, \bibinfo {author} {\bibfnamefont {J.}~\bibnamefont {Skotheim}},
  \bibinfo {author} {\bibfnamefont {L.}~\bibnamefont {Mahadevan}},\ and\
  \bibinfo {author} {\bibfnamefont {D.~A.}\ \bibnamefont {Weitz}},\ }\bibfield
  {title} {\bibinfo {title} {Gravitational collapse of colloidal gels},\
  }\href@noop {} {\bibfield  {journal} {\bibinfo  {journal} {Physical review
  letters}\ }\textbf {\bibinfo {volume} {94}},\ \bibinfo {pages} {218302}
  (\bibinfo {year} {2005})}\BibitemShut {NoStop}%
\bibitem [{\citenamefont {Derec}\ \emph {et~al.}(2003)\citenamefont {Derec},
  \citenamefont {Senis}, \citenamefont {Talini},\ and\ \citenamefont
  {Allain}}]{derec2003rapid}%
  \BibitemOpen
  \bibfield  {author} {\bibinfo {author} {\bibfnamefont {C.}~\bibnamefont
  {Derec}}, \bibinfo {author} {\bibfnamefont {D.}~\bibnamefont {Senis}},
  \bibinfo {author} {\bibfnamefont {L.}~\bibnamefont {Talini}},\ and\ \bibinfo
  {author} {\bibfnamefont {C.}~\bibnamefont {Allain}},\ }\bibfield  {title}
  {\bibinfo {title} {Rapid settling of a colloidal gel},\ }\href@noop {}
  {\bibfield  {journal} {\bibinfo  {journal} {Physical Review E}\ }\textbf
  {\bibinfo {volume} {67}},\ \bibinfo {pages} {062401} (\bibinfo {year}
  {2003})}\BibitemShut {NoStop}%
\bibitem [{\citenamefont {Allain}\ \emph {et~al.}(1995)\citenamefont {Allain},
  \citenamefont {Cloitre},\ and\ \citenamefont
  {Wafra}}]{allain1995aggregation}%
  \BibitemOpen
  \bibfield  {author} {\bibinfo {author} {\bibfnamefont {C.}~\bibnamefont
  {Allain}}, \bibinfo {author} {\bibfnamefont {M.}~\bibnamefont {Cloitre}},\
  and\ \bibinfo {author} {\bibfnamefont {M.}~\bibnamefont {Wafra}},\ }\bibfield
   {title} {\bibinfo {title} {Aggregation and sedimentation in colloidal
  suspensions},\ }\href@noop {} {\bibfield  {journal} {\bibinfo  {journal}
  {Physical review letters}\ }\textbf {\bibinfo {volume} {74}},\ \bibinfo
  {pages} {1478} (\bibinfo {year} {1995})}\BibitemShut {NoStop}%
\bibitem [{\citenamefont {Darras}\ \emph
  {et~al.}(2021{\natexlab{a}})\citenamefont {Darras}, \citenamefont {Dasanna},
  \citenamefont {John}, \citenamefont {Gompper}, \citenamefont {Kaestner},
  \citenamefont {Fedosov},\ and\ \citenamefont {Wagner}}]{JointLetter}%
  \BibitemOpen
  \bibfield  {author} {\bibinfo {author} {\bibfnamefont {A.}~\bibnamefont
  {Darras}}, \bibinfo {author} {\bibfnamefont {A.~K.}\ \bibnamefont {Dasanna}},
  \bibinfo {author} {\bibfnamefont {T.}~\bibnamefont {John}}, \bibinfo {author}
  {\bibfnamefont {G.}~\bibnamefont {Gompper}}, \bibinfo {author} {\bibfnamefont
  {L.}~\bibnamefont {Kaestner}}, \bibinfo {author} {\bibfnamefont {D.~A.}\
  \bibnamefont {Fedosov}},\ and\ \bibinfo {author} {\bibfnamefont
  {C.}~\bibnamefont {Wagner}},\ }\bibfield  {title} {\bibinfo {title}
  {Erythrocyte sedimentation: fracture and collapse of a high-volume-fraction
  soft-colloid gel},\ }\href@noop {} {\bibfield  {journal} {\bibinfo  {journal}
  {Physical Review Letters}\ }\textbf {\bibinfo {volume} {submitted}} (\bibinfo
  {year} {2021}{\natexlab{a}})}\BibitemShut {NoStop}%
\bibitem [{\citenamefont {Pribush}\ \emph {et~al.}(2010)\citenamefont
  {Pribush}, \citenamefont {Meyerstein},\ and\ \citenamefont
  {Meyerstein}}]{pribush2010mechanism1}%
  \BibitemOpen
  \bibfield  {author} {\bibinfo {author} {\bibfnamefont {A.}~\bibnamefont
  {Pribush}}, \bibinfo {author} {\bibfnamefont {D.}~\bibnamefont
  {Meyerstein}},\ and\ \bibinfo {author} {\bibfnamefont {N.}~\bibnamefont
  {Meyerstein}},\ }\bibfield  {title} {\bibinfo {title} {The mechanism of
  erythrocyte sedimentation. part 1: Channeling in sedimenting blood},\
  }\href@noop {} {\bibfield  {journal} {\bibinfo  {journal} {Colloids and
  surfaces B: Biointerfaces}\ }\textbf {\bibinfo {volume} {75}},\ \bibinfo
  {pages} {214} (\bibinfo {year} {2010})}\BibitemShut {NoStop}%
\bibitem [{\citenamefont {Channell}\ \emph {et~al.}(2000)\citenamefont
  {Channell}, \citenamefont {Miller},\ and\ \citenamefont
  {Zukoski}}]{channell2000effects}%
  \BibitemOpen
  \bibfield  {author} {\bibinfo {author} {\bibfnamefont {G.~M.}\ \bibnamefont
  {Channell}}, \bibinfo {author} {\bibfnamefont {K.~T.}\ \bibnamefont
  {Miller}},\ and\ \bibinfo {author} {\bibfnamefont {C.~F.}\ \bibnamefont
  {Zukoski}},\ }\bibfield  {title} {\bibinfo {title} {Effects of microstructure
  on the compressive yield stress},\ }\href@noop {} {\bibfield  {journal}
  {\bibinfo  {journal} {AIChE journal}\ }\textbf {\bibinfo {volume} {46}},\
  \bibinfo {pages} {72} (\bibinfo {year} {2000})}\BibitemShut {NoStop}%
\bibitem [{\citenamefont {Kilfoil}\ \emph {et~al.}(2003)\citenamefont
  {Kilfoil}, \citenamefont {Pashkovski}, \citenamefont {Masters},\ and\
  \citenamefont {Weitz}}]{kilfoil2003dynamics}%
  \BibitemOpen
  \bibfield  {author} {\bibinfo {author} {\bibfnamefont {M.~L.}\ \bibnamefont
  {Kilfoil}}, \bibinfo {author} {\bibfnamefont {E.~E.}\ \bibnamefont
  {Pashkovski}}, \bibinfo {author} {\bibfnamefont {J.~A.}\ \bibnamefont
  {Masters}},\ and\ \bibinfo {author} {\bibfnamefont {D.}~\bibnamefont
  {Weitz}},\ }\bibfield  {title} {\bibinfo {title} {Dynamics of weakly
  aggregated colloidal particles},\ }\href@noop {} {\bibfield  {journal}
  {\bibinfo  {journal} {Philosophical Transactions of the Royal Society of
  London. Series A: Mathematical, Physical and Engineering Sciences}\ }\textbf
  {\bibinfo {volume} {361}},\ \bibinfo {pages} {753} (\bibinfo {year}
  {2003})}\BibitemShut {NoStop}%
\bibitem [{\citenamefont {Kamp}\ and\ \citenamefont
  {Kilfoil}(2009)}]{kamp2009universal}%
  \BibitemOpen
  \bibfield  {author} {\bibinfo {author} {\bibfnamefont {S.~W.}\ \bibnamefont
  {Kamp}}\ and\ \bibinfo {author} {\bibfnamefont {M.~L.}\ \bibnamefont
  {Kilfoil}},\ }\bibfield  {title} {\bibinfo {title} {Universal behaviour in
  the mechanical properties of weakly aggregated colloidal particles},\
  }\href@noop {} {\bibfield  {journal} {\bibinfo  {journal} {Soft matter}\
  }\textbf {\bibinfo {volume} {5}},\ \bibinfo {pages} {2438} (\bibinfo {year}
  {2009})}\BibitemShut {NoStop}%
\bibitem [{\citenamefont {Dinsmore}\ \emph {et~al.}(2006)\citenamefont
  {Dinsmore}, \citenamefont {Prasad}, \citenamefont {Wong},\ and\ \citenamefont
  {Weitz}}]{dinsmore2006microscopic}%
  \BibitemOpen
  \bibfield  {author} {\bibinfo {author} {\bibfnamefont {A.}~\bibnamefont
  {Dinsmore}}, \bibinfo {author} {\bibfnamefont {V.}~\bibnamefont {Prasad}},
  \bibinfo {author} {\bibfnamefont {I.}~\bibnamefont {Wong}},\ and\ \bibinfo
  {author} {\bibfnamefont {D.}~\bibnamefont {Weitz}},\ }\bibfield  {title}
  {\bibinfo {title} {Microscopic structure and elasticity of weakly aggregated
  colloidal gels},\ }\href@noop {} {\bibfield  {journal} {\bibinfo  {journal}
  {Physical review letters}\ }\textbf {\bibinfo {volume} {96}},\ \bibinfo
  {pages} {185502} (\bibinfo {year} {2006})}\BibitemShut {NoStop}%
\bibitem [{\citenamefont {Lindstr{\"o}m}\ \emph {et~al.}(2012)\citenamefont
  {Lindstr{\"o}m}, \citenamefont {Kodger}, \citenamefont {Sprakel},\ and\
  \citenamefont {Weitz}}]{lindstrom2012}%
  \BibitemOpen
  \bibfield  {author} {\bibinfo {author} {\bibfnamefont {S.~B.}\ \bibnamefont
  {Lindstr{\"o}m}}, \bibinfo {author} {\bibfnamefont {T.~E.}\ \bibnamefont
  {Kodger}}, \bibinfo {author} {\bibfnamefont {J.}~\bibnamefont {Sprakel}},\
  and\ \bibinfo {author} {\bibfnamefont {D.~A.}\ \bibnamefont {Weitz}},\
  }\bibfield  {title} {\bibinfo {title} {Structures, stresses, and fluctuations
  in the delayed failure of colloidal gels},\ }\href@noop {} {\bibfield
  {journal} {\bibinfo  {journal} {Soft Matter}\ }\textbf {\bibinfo {volume}
  {8}},\ \bibinfo {pages} {3657} (\bibinfo {year} {2012})}\BibitemShut
  {NoStop}%
\bibitem [{\citenamefont {Bedell}\ and\ \citenamefont
  {Bush}(1985)}]{bedell1985}%
  \BibitemOpen
  \bibfield  {author} {\bibinfo {author} {\bibfnamefont {S.~E.}\ \bibnamefont
  {Bedell}}\ and\ \bibinfo {author} {\bibfnamefont {B.~T.}\ \bibnamefont
  {Bush}},\ }\bibfield  {title} {\bibinfo {title} {{Erythrocyte sedimentation
  rate. From folklore to facts}},\ }\href@noop {} {\bibfield  {journal}
  {\bibinfo  {journal} {The American journal of medicine}\ }\textbf {\bibinfo
  {volume} {78}},\ \bibinfo {pages} {1001} (\bibinfo {year}
  {1985})}\BibitemShut {NoStop}%
\bibitem [{\citenamefont {Kratz}\ \emph {et~al.}(2017)\citenamefont {Kratz},
  \citenamefont {Plebani}, \citenamefont {Peng}, \citenamefont {Lee},
  \citenamefont {McCafferty}, \citenamefont {Machin},\ and\ \citenamefont {for
  Standardization~in Haematology~(ICSH)}}]{kratz2017}%
  \BibitemOpen
  \bibfield  {author} {\bibinfo {author} {\bibfnamefont {A.}~\bibnamefont
  {Kratz}}, \bibinfo {author} {\bibfnamefont {M.}~\bibnamefont {Plebani}},
  \bibinfo {author} {\bibfnamefont {M.}~\bibnamefont {Peng}}, \bibinfo {author}
  {\bibfnamefont {Y.}~\bibnamefont {Lee}}, \bibinfo {author} {\bibfnamefont
  {R.}~\bibnamefont {McCafferty}}, \bibinfo {author} {\bibfnamefont
  {S.}~\bibnamefont {Machin}},\ and\ \bibinfo {author} {\bibfnamefont {I.~C.}\
  \bibnamefont {for Standardization~in Haematology~(ICSH)}},\ }\bibfield
  {title} {\bibinfo {title} {{ICSH recommendations for modified and alternate
  methods measuring the erythrocyte sedimentation rate}},\ }\href@noop {}
  {\bibfield  {journal} {\bibinfo  {journal} {International journal of
  laboratory hematology}\ }\textbf {\bibinfo {volume} {39}},\ \bibinfo {pages}
  {448} (\bibinfo {year} {2017})}\BibitemShut {NoStop}%
\bibitem [{\citenamefont {Terzaghi}\ \emph {et~al.}(1925)\citenamefont
  {Terzaghi} \emph {et~al.}}]{terzaghi1925}%
  \BibitemOpen
  \bibfield  {author} {\bibinfo {author} {\bibfnamefont {K.}~\bibnamefont
  {Terzaghi}} \emph {et~al.},\ }\href@noop {} {\emph {\bibinfo {title}
  {Erdbaumechanik auf bodenphysikalischer Grundlage}}},\ \bibinfo {type} {Tech.
  Rep.}\ (\bibinfo {year} {1925})\BibitemShut {NoStop}%
\bibitem [{\citenamefont {Carman}(1939)}]{carman1939permeability}%
  \BibitemOpen
  \bibfield  {author} {\bibinfo {author} {\bibfnamefont {P.~C.}\ \bibnamefont
  {Carman}},\ }\bibfield  {title} {\bibinfo {title} {Permeability of saturated
  sands, soils and clays},\ }\href@noop {} {\bibfield  {journal} {\bibinfo
  {journal} {The Journal of Agricultural Science}\ }\textbf {\bibinfo {volume}
  {29}},\ \bibinfo {pages} {262} (\bibinfo {year} {1939})}\BibitemShut
  {NoStop}%
\bibitem [{\citenamefont {Ozgumus}\ \emph {et~al.}(2014)\citenamefont
  {Ozgumus}, \citenamefont {Mobedi},\ and\ \citenamefont
  {Ozkol}}]{ozgumus2014determination}%
  \BibitemOpen
  \bibfield  {author} {\bibinfo {author} {\bibfnamefont {T.}~\bibnamefont
  {Ozgumus}}, \bibinfo {author} {\bibfnamefont {M.}~\bibnamefont {Mobedi}},\
  and\ \bibinfo {author} {\bibfnamefont {U.}~\bibnamefont {Ozkol}},\ }\bibfield
   {title} {\bibinfo {title} {Determination of kozeny constant based on
  porosity and pore to throat size ratio in porous medium with rectangular
  rods},\ }\href@noop {} {\bibfield  {journal} {\bibinfo  {journal}
  {Engineering Applications of Computational Fluid Mechanics}\ }\textbf
  {\bibinfo {volume} {8}},\ \bibinfo {pages} {308} (\bibinfo {year}
  {2014})}\BibitemShut {NoStop}%
\bibitem [{\citenamefont {Heijs}\ and\ \citenamefont
  {Lowe}(1995)}]{heijs1995numerical}%
  \BibitemOpen
  \bibfield  {author} {\bibinfo {author} {\bibfnamefont {A.~W.}\ \bibnamefont
  {Heijs}}\ and\ \bibinfo {author} {\bibfnamefont {C.~P.}\ \bibnamefont
  {Lowe}},\ }\bibfield  {title} {\bibinfo {title} {Numerical evaluation of the
  permeability and the kozeny constant for two types of porous media},\
  }\href@noop {} {\bibfield  {journal} {\bibinfo  {journal} {Physical Review
  E}\ }\textbf {\bibinfo {volume} {51}},\ \bibinfo {pages} {4346} (\bibinfo
  {year} {1995})}\BibitemShut {NoStop}%
\bibitem [{\citenamefont {Xu}\ and\ \citenamefont
  {Yu}(2008)}]{xu2008developing}%
  \BibitemOpen
  \bibfield  {author} {\bibinfo {author} {\bibfnamefont {P.}~\bibnamefont
  {Xu}}\ and\ \bibinfo {author} {\bibfnamefont {B.}~\bibnamefont {Yu}},\
  }\bibfield  {title} {\bibinfo {title} {Developing a new form of permeability
  and kozeny--carman constant for homogeneous porous media by means of fractal
  geometry},\ }\href@noop {} {\bibfield  {journal} {\bibinfo  {journal}
  {Advances in water resources}\ }\textbf {\bibinfo {volume} {31}},\ \bibinfo
  {pages} {74} (\bibinfo {year} {2008})}\BibitemShut {NoStop}%
\bibitem [{\citenamefont {Norouzi}\ \emph {et~al.}(2017)\citenamefont
  {Norouzi}, \citenamefont {Bhakta},\ and\ \citenamefont
  {Grover}}]{norouzi2017sorting}%
  \BibitemOpen
  \bibfield  {author} {\bibinfo {author} {\bibfnamefont {N.}~\bibnamefont
  {Norouzi}}, \bibinfo {author} {\bibfnamefont {H.~C.}\ \bibnamefont
  {Bhakta}},\ and\ \bibinfo {author} {\bibfnamefont {W.~H.}\ \bibnamefont
  {Grover}},\ }\bibfield  {title} {\bibinfo {title} {Sorting cells by their
  density},\ }\href@noop {} {\bibfield  {journal} {\bibinfo  {journal} {PloS
  one}\ }\textbf {\bibinfo {volume} {12}},\ \bibinfo {pages} {e0180520}
  (\bibinfo {year} {2017})}\BibitemShut {NoStop}%
\bibitem [{\citenamefont {Trudnowski}\ and\ \citenamefont
  {Rico}(1974)}]{trudnowski1974specific}%
  \BibitemOpen
  \bibfield  {author} {\bibinfo {author} {\bibfnamefont {R.~J.}\ \bibnamefont
  {Trudnowski}}\ and\ \bibinfo {author} {\bibfnamefont {R.~C.}\ \bibnamefont
  {Rico}},\ }\bibfield  {title} {\bibinfo {title} {Specific gravity of blood
  and plasma at 4 and 37 c},\ }\href@noop {} {\bibfield  {journal} {\bibinfo
  {journal} {Clinical chemistry}\ }\textbf {\bibinfo {volume} {20}},\ \bibinfo
  {pages} {615} (\bibinfo {year} {1974})}\BibitemShut {NoStop}%
\bibitem [{\citenamefont {Brust}\ \emph {et~al.}(2014)\citenamefont {Brust},
  \citenamefont {Aouane}, \citenamefont {Thi{\'e}baud}, \citenamefont
  {Flormann}, \citenamefont {Verdier}, \citenamefont {Kaestner}, \citenamefont
  {Laschke}, \citenamefont {Selmi}, \citenamefont {Benyoussef}, \citenamefont
  {Podgorski} \emph {et~al.}}]{brust2014plasma}%
  \BibitemOpen
  \bibfield  {author} {\bibinfo {author} {\bibfnamefont {M.}~\bibnamefont
  {Brust}}, \bibinfo {author} {\bibfnamefont {O.}~\bibnamefont {Aouane}},
  \bibinfo {author} {\bibfnamefont {M.}~\bibnamefont {Thi{\'e}baud}}, \bibinfo
  {author} {\bibfnamefont {D.}~\bibnamefont {Flormann}}, \bibinfo {author}
  {\bibfnamefont {C.}~\bibnamefont {Verdier}}, \bibinfo {author} {\bibfnamefont
  {L.}~\bibnamefont {Kaestner}}, \bibinfo {author} {\bibfnamefont
  {M.}~\bibnamefont {Laschke}}, \bibinfo {author} {\bibfnamefont
  {H.}~\bibnamefont {Selmi}}, \bibinfo {author} {\bibfnamefont
  {A.}~\bibnamefont {Benyoussef}}, \bibinfo {author} {\bibfnamefont
  {T.}~\bibnamefont {Podgorski}}, \emph {et~al.},\ }\bibfield  {title}
  {\bibinfo {title} {The plasma protein fibrinogen stabilizes clusters of red
  blood cells in microcapillary flows},\ }\href@noop {} {\bibfield  {journal}
  {\bibinfo  {journal} {Scientific reports}\ }\textbf {\bibinfo {volume} {4}},\
  \bibinfo {pages} {1} (\bibinfo {year} {2014})}\BibitemShut {NoStop}%
\bibitem [{\citenamefont {Flormann}\ \emph {et~al.}(2015)\citenamefont
  {Flormann}, \citenamefont {Kuder}, \citenamefont {Lipp}, \citenamefont
  {Wagner},\ and\ \citenamefont {Kaestner}}]{flormann2015there}%
  \BibitemOpen
  \bibfield  {author} {\bibinfo {author} {\bibfnamefont {D.}~\bibnamefont
  {Flormann}}, \bibinfo {author} {\bibfnamefont {E.}~\bibnamefont {Kuder}},
  \bibinfo {author} {\bibfnamefont {P.}~\bibnamefont {Lipp}}, \bibinfo {author}
  {\bibfnamefont {C.}~\bibnamefont {Wagner}},\ and\ \bibinfo {author}
  {\bibfnamefont {L.}~\bibnamefont {Kaestner}},\ }\bibfield  {title} {\bibinfo
  {title} {Is there a role of c-reactive protein in red blood cell
  aggregation?},\ }\href@noop {} {\bibfield  {journal} {\bibinfo  {journal}
  {International journal of laboratory hematology}\ }\textbf {\bibinfo {volume}
  {37}},\ \bibinfo {pages} {474} (\bibinfo {year} {2015})}\BibitemShut
  {NoStop}%
\bibitem [{\citenamefont {Lee}\ \emph {et~al.}(2016)\citenamefont {Lee},
  \citenamefont {Kinnunen}, \citenamefont {Khokhlova}, \citenamefont {Lyubin},
  \citenamefont {Priezzhev}, \citenamefont {Meglinski},\ and\ \citenamefont
  {Fedyanin}}]{lee2016optical}%
  \BibitemOpen
  \bibfield  {author} {\bibinfo {author} {\bibfnamefont {K.}~\bibnamefont
  {Lee}}, \bibinfo {author} {\bibfnamefont {M.}~\bibnamefont {Kinnunen}},
  \bibinfo {author} {\bibfnamefont {M.~D.}\ \bibnamefont {Khokhlova}}, \bibinfo
  {author} {\bibfnamefont {E.~V.}\ \bibnamefont {Lyubin}}, \bibinfo {author}
  {\bibfnamefont {A.~V.}\ \bibnamefont {Priezzhev}}, \bibinfo {author}
  {\bibfnamefont {I.}~\bibnamefont {Meglinski}},\ and\ \bibinfo {author}
  {\bibfnamefont {A.~A.}\ \bibnamefont {Fedyanin}},\ }\bibfield  {title}
  {\bibinfo {title} {Optical tweezers study of red blood cell aggregation and
  disaggregation in plasma and protein solutions},\ }\href@noop {} {\bibfield
  {journal} {\bibinfo  {journal} {Journal of biomedical optics}\ }\textbf
  {\bibinfo {volume} {21}},\ \bibinfo {pages} {035001} (\bibinfo {year}
  {2016})}\BibitemShut {NoStop}%
\bibitem [{\citenamefont {Issaq}\ \emph {et~al.}(2007)\citenamefont {Issaq},
  \citenamefont {Xiao},\ and\ \citenamefont {Veenstra}}]{issaq2007serum}%
  \BibitemOpen
  \bibfield  {author} {\bibinfo {author} {\bibfnamefont {H.~J.}\ \bibnamefont
  {Issaq}}, \bibinfo {author} {\bibfnamefont {Z.}~\bibnamefont {Xiao}},\ and\
  \bibinfo {author} {\bibfnamefont {T.~D.}\ \bibnamefont {Veenstra}},\
  }\bibfield  {title} {\bibinfo {title} {Serum and plasma proteomics},\
  }\href@noop {} {\bibfield  {journal} {\bibinfo  {journal} {Chemical reviews}\
  }\textbf {\bibinfo {volume} {107}},\ \bibinfo {pages} {3601} (\bibinfo {year}
  {2007})}\BibitemShut {NoStop}%
\bibitem [{\citenamefont {Yu}\ \emph {et~al.}(2011)\citenamefont {Yu},
  \citenamefont {Kastenm{\"u}ller}, \citenamefont {He}, \citenamefont
  {Belcredi}, \citenamefont {M{\"o}ller}, \citenamefont {Prehn}, \citenamefont
  {Mendes}, \citenamefont {Wahl}, \citenamefont {Roemisch-Margl}, \citenamefont
  {Ceglarek} \emph {et~al.}}]{yu2011differences}%
  \BibitemOpen
  \bibfield  {author} {\bibinfo {author} {\bibfnamefont {Z.}~\bibnamefont
  {Yu}}, \bibinfo {author} {\bibfnamefont {G.}~\bibnamefont
  {Kastenm{\"u}ller}}, \bibinfo {author} {\bibfnamefont {Y.}~\bibnamefont
  {He}}, \bibinfo {author} {\bibfnamefont {P.}~\bibnamefont {Belcredi}},
  \bibinfo {author} {\bibfnamefont {G.}~\bibnamefont {M{\"o}ller}}, \bibinfo
  {author} {\bibfnamefont {C.}~\bibnamefont {Prehn}}, \bibinfo {author}
  {\bibfnamefont {J.}~\bibnamefont {Mendes}}, \bibinfo {author} {\bibfnamefont
  {S.}~\bibnamefont {Wahl}}, \bibinfo {author} {\bibfnamefont {W.}~\bibnamefont
  {Roemisch-Margl}}, \bibinfo {author} {\bibfnamefont {U.}~\bibnamefont
  {Ceglarek}}, \emph {et~al.},\ }\bibfield  {title} {\bibinfo {title}
  {Differences between human plasma and serum metabolite profiles},\
  }\href@noop {} {\bibfield  {journal} {\bibinfo  {journal} {PloS one}\
  }\textbf {\bibinfo {volume} {6}},\ \bibinfo {pages} {e21230} (\bibinfo {year}
  {2011})}\BibitemShut {NoStop}%
\bibitem [{\citenamefont {Stauffer}\ and\ \citenamefont
  {Aharony}(2018)}]{stauffer2018introduction}%
  \BibitemOpen
  \bibfield  {author} {\bibinfo {author} {\bibfnamefont {D.}~\bibnamefont
  {Stauffer}}\ and\ \bibinfo {author} {\bibfnamefont {A.}~\bibnamefont
  {Aharony}},\ }\href@noop {} {\emph {\bibinfo {title} {Introduction to
  percolation theory}}}\ (\bibinfo  {publisher} {CRC press},\ \bibinfo {year}
  {2018})\BibitemShut {NoStop}%
\bibitem [{\citenamefont {Noguchi}\ and\ \citenamefont
  {Gompper}(2005)}]{Noguchi_STV_2005}%
  \BibitemOpen
  \bibfield  {author} {\bibinfo {author} {\bibfnamefont {H.}~\bibnamefont
  {Noguchi}}\ and\ \bibinfo {author} {\bibfnamefont {G.}~\bibnamefont
  {Gompper}},\ }\bibfield  {title} {\bibinfo {title} {Shape transitions of
  fluid vesicles and red blood cells in capillary flows},\ }\href@noop {}
  {\bibfield  {journal} {\bibinfo  {journal} {Proc. Natl. Acad. Sci. USA}\
  }\textbf {\bibinfo {volume} {102}},\ \bibinfo {pages} {14159} (\bibinfo
  {year} {2005})}\BibitemShut {NoStop}%
\bibitem [{\citenamefont {Fedosov}\ \emph
  {et~al.}(2010{\natexlab{a}})\citenamefont {Fedosov}, \citenamefont
  {Caswell},\ and\ \citenamefont {Karniadakis}}]{Fedosov_RBC_2010}%
  \BibitemOpen
  \bibfield  {author} {\bibinfo {author} {\bibfnamefont {D.~A.}\ \bibnamefont
  {Fedosov}}, \bibinfo {author} {\bibfnamefont {B.}~\bibnamefont {Caswell}},\
  and\ \bibinfo {author} {\bibfnamefont {G.~E.}\ \bibnamefont {Karniadakis}},\
  }\bibfield  {title} {\bibinfo {title} {A multiscale red blood cell model with
  accurate mechanics, rheology, and dynamics},\ }\href@noop {} {\bibfield
  {journal} {\bibinfo  {journal} {Biophys. J.}\ }\textbf {\bibinfo {volume}
  {98}},\ \bibinfo {pages} {2215} (\bibinfo {year}
  {2010}{\natexlab{a}})}\BibitemShut {NoStop}%
\bibitem [{\citenamefont {Fedosov}\ \emph
  {et~al.}(2010{\natexlab{b}})\citenamefont {Fedosov}, \citenamefont
  {Caswell},\ and\ \citenamefont {Karniadakis}}]{Fedosov_SCG_2010}%
  \BibitemOpen
  \bibfield  {author} {\bibinfo {author} {\bibfnamefont {D.~A.}\ \bibnamefont
  {Fedosov}}, \bibinfo {author} {\bibfnamefont {B.}~\bibnamefont {Caswell}},\
  and\ \bibinfo {author} {\bibfnamefont {G.~E.}\ \bibnamefont {Karniadakis}},\
  }\bibfield  {title} {\bibinfo {title} {Systematic coarse-graining of
  spectrin-level red blood cell models},\ }\href@noop {} {\bibfield  {journal}
  {\bibinfo  {journal} {Comput. Meth. Appl. Mech. Eng.}\ }\textbf {\bibinfo
  {volume} {199}},\ \bibinfo {pages} {1937} (\bibinfo {year}
  {2010}{\natexlab{b}})}\BibitemShut {NoStop}%
\bibitem [{\citenamefont {Fedosov}\ \emph {et~al.}(2011)\citenamefont
  {Fedosov}, \citenamefont {Pan}, \citenamefont {Caswell}, \citenamefont
  {Gompper},\ and\ \citenamefont {Karniadakis}}]{Fedosov_PBV_2011}%
  \BibitemOpen
  \bibfield  {author} {\bibinfo {author} {\bibfnamefont {D.~A.}\ \bibnamefont
  {Fedosov}}, \bibinfo {author} {\bibfnamefont {W.}~\bibnamefont {Pan}},
  \bibinfo {author} {\bibfnamefont {B.}~\bibnamefont {Caswell}}, \bibinfo
  {author} {\bibfnamefont {G.}~\bibnamefont {Gompper}},\ and\ \bibinfo {author}
  {\bibfnamefont {G.~E.}\ \bibnamefont {Karniadakis}},\ }\bibfield  {title}
  {\bibinfo {title} {Predicting human blood viscosity in silico},\ }\href@noop
  {} {\bibfield  {journal} {\bibinfo  {journal} {Proc. Natl. Acad. Sci. USA}\
  }\textbf {\bibinfo {volume} {108}},\ \bibinfo {pages} {11772} (\bibinfo
  {year} {2011})}\BibitemShut {NoStop}%
\bibitem [{\citenamefont {Ermolinskiy}\ \emph {et~al.}(2020)\citenamefont
  {Ermolinskiy}, \citenamefont {Lugovtsov}, \citenamefont {Yaya}, \citenamefont
  {Lee}, \citenamefont {Kaestner}, \citenamefont {Wagner},\ and\ \citenamefont
  {Priezzhev}}]{ermolinskiy2020effect}%
  \BibitemOpen
  \bibfield  {author} {\bibinfo {author} {\bibfnamefont {P.}~\bibnamefont
  {Ermolinskiy}}, \bibinfo {author} {\bibfnamefont {A.}~\bibnamefont
  {Lugovtsov}}, \bibinfo {author} {\bibfnamefont {F.}~\bibnamefont {Yaya}},
  \bibinfo {author} {\bibfnamefont {K.}~\bibnamefont {Lee}}, \bibinfo {author}
  {\bibfnamefont {L.}~\bibnamefont {Kaestner}}, \bibinfo {author}
  {\bibfnamefont {C.}~\bibnamefont {Wagner}},\ and\ \bibinfo {author}
  {\bibfnamefont {A.}~\bibnamefont {Priezzhev}},\ }\bibfield  {title} {\bibinfo
  {title} {Effect of red blood cell aging in vivo on their aggregation
  properties in vitro: Measurements with laser tweezers},\ }\href@noop {}
  {\bibfield  {journal} {\bibinfo  {journal} {Applied Sciences}\ }\textbf
  {\bibinfo {volume} {10}},\ \bibinfo {pages} {7581} (\bibinfo {year}
  {2020})}\BibitemShut {NoStop}%
\bibitem [{\citenamefont {Espa\~{n}ol}\ and\ \citenamefont
  {Revenga}(2003)}]{Espanol_SDPD_2003}%
  \BibitemOpen
  \bibfield  {author} {\bibinfo {author} {\bibfnamefont {P.}~\bibnamefont
  {Espa\~{n}ol}}\ and\ \bibinfo {author} {\bibfnamefont {M.}~\bibnamefont
  {Revenga}},\ }\bibfield  {title} {\bibinfo {title} {Smoothed dissipative
  particle dynamics},\ }\href@noop {} {\bibfield  {journal} {\bibinfo
  {journal} {Phys. Rev. E}\ }\textbf {\bibinfo {volume} {67}},\ \bibinfo
  {pages} {026705} (\bibinfo {year} {2003})}\BibitemShut {NoStop}%
\bibitem [{\citenamefont {M{\"u}ller}\ \emph {et~al.}(2015)\citenamefont
  {M{\"u}ller}, \citenamefont {Fedosov},\ and\ \citenamefont
  {Gompper}}]{Mueller_SDPD_2015}%
  \BibitemOpen
  \bibfield  {author} {\bibinfo {author} {\bibfnamefont {K.}~\bibnamefont
  {M{\"u}ller}}, \bibinfo {author} {\bibfnamefont {D.~A.}\ \bibnamefont
  {Fedosov}},\ and\ \bibinfo {author} {\bibfnamefont {G.}~\bibnamefont
  {Gompper}},\ }\bibfield  {title} {\bibinfo {title} {Smoothed dissipative
  particle dynamics with angular momentum conservation},\ }\href@noop {}
  {\bibfield  {journal} {\bibinfo  {journal} {J. Comp. Phys.}\ }\textbf
  {\bibinfo {volume} {281}},\ \bibinfo {pages} {301} (\bibinfo {year}
  {2015})}\BibitemShut {NoStop}%
\bibitem [{\citenamefont {Evans}\ and\ \citenamefont
  {Skalak}(1980)}]{Evans_MTB_1980}%
  \BibitemOpen
  \bibfield  {author} {\bibinfo {author} {\bibfnamefont {E.~A.}\ \bibnamefont
  {Evans}}\ and\ \bibinfo {author} {\bibfnamefont {R.}~\bibnamefont {Skalak}},\
  }\href@noop {} {\emph {\bibinfo {title} {Mechanics and thermodynamics of
  biomembranes}}}\ (\bibinfo  {publisher} {CRC Press, Inc.},\ \bibinfo
  {address} {Boca Raton, Florida},\ \bibinfo {year} {1980})\BibitemShut
  {NoStop}%
\bibitem [{\citenamefont {Evans}(1983)}]{Evans_BEM_1983}%
  \BibitemOpen
  \bibfield  {author} {\bibinfo {author} {\bibfnamefont {E.~A.}\ \bibnamefont
  {Evans}},\ }\bibfield  {title} {\bibinfo {title} {Bending elastic modulus of
  red blood cell membrane derived from buckling instability in micropipet
  aspiration tests},\ }\href@noop {} {\bibfield  {journal} {\bibinfo  {journal}
  {Biophys. J.}\ }\textbf {\bibinfo {volume} {43}},\ \bibinfo {pages} {27}
  (\bibinfo {year} {1983})}\BibitemShut {NoStop}%
\bibitem [{\citenamefont {Dao}\ \emph {et~al.}(2003)\citenamefont {Dao},
  \citenamefont {Lim},\ and\ \citenamefont {Suresh}}]{Dao_RBC_2003}%
  \BibitemOpen
  \bibfield  {author} {\bibinfo {author} {\bibfnamefont {M.}~\bibnamefont
  {Dao}}, \bibinfo {author} {\bibfnamefont {C.~T.}\ \bibnamefont {Lim}},\ and\
  \bibinfo {author} {\bibfnamefont {S.}~\bibnamefont {Suresh}},\ }\bibfield
  {title} {\bibinfo {title} {Mechanics of the human red blood cell deformed by
  optical tweezers},\ }\href@noop {} {\bibfield  {journal} {\bibinfo  {journal}
  {J. Mech. Phys. Solids}\ }\textbf {\bibinfo {volume} {51}},\ \bibinfo {pages}
  {2259} (\bibinfo {year} {2003})}\BibitemShut {NoStop}%
\bibitem [{\citenamefont {Peltom\"aki}\ and\ \citenamefont
  {Gompper}(2013)}]{peltomaeki2013}%
  \BibitemOpen
  \bibfield  {author} {\bibinfo {author} {\bibfnamefont {M.}~\bibnamefont
  {Peltom\"aki}}\ and\ \bibinfo {author} {\bibfnamefont {G.}~\bibnamefont
  {Gompper}},\ }\bibfield  {title} {\bibinfo {title} {Sedimentation of single
  red blood cells},\ }\href {https://doi.org/10.1039/C3SM50592H} {\bibfield
  {journal} {\bibinfo  {journal} {Soft Matter}\ }\textbf {\bibinfo {volume}
  {9}},\ \bibinfo {pages} {8346} (\bibinfo {year} {2013})}\BibitemShut
  {NoStop}%
\bibitem [{Sup()}]{SuppMat}%
  \BibitemOpen
  \href@noop {} {\bibinfo {title} {{See Supplemental Material at [URL will be
  inserted by publisher] for Supplemental Materials. They contain: Supplemental
  Figure 1 (illustration of pore sizes determination by numerical simulations),
  Supplemental Figure 2 (Holes size PDF from simulations), and Supplemental
  Movie S1/H40ImageJ.avi (Illustrative movie generated from numerical
  simulations).}}}\BibitemShut {Stop}%
\bibitem [{\citenamefont {Russel}\ \emph {et~al.}(1989)\citenamefont {Russel},
  \citenamefont {Saville},\ and\ \citenamefont
  {Schowalter}}]{russel1989colloidal}%
  \BibitemOpen
  \bibfield  {author} {\bibinfo {author} {\bibfnamefont {W.}~\bibnamefont
  {Russel}}, \bibinfo {author} {\bibfnamefont {D.}~\bibnamefont {Saville}},\
  and\ \bibinfo {author} {\bibfnamefont {W.}~\bibnamefont {Schowalter}},\
  }\href@noop {} {\bibinfo {title} {Colloidal dispersions cambridge univ}}
  (\bibinfo {year} {1989})\BibitemShut {NoStop}%
\bibitem [{\citenamefont {Darras}\ \emph
  {et~al.}(2021{\natexlab{b}})\citenamefont {Darras}, \citenamefont {Peikert},
  \citenamefont {Rabe}, \citenamefont {Yaya}, \citenamefont {Simionato},
  \citenamefont {John}, \citenamefont {Dasanna}, \citenamefont {Buvalyy},
  \citenamefont {Geisel}, \citenamefont {Hermann} \emph
  {et~al.}}]{Darras2020Erythrocytes}%
  \BibitemOpen
  \bibfield  {author} {\bibinfo {author} {\bibfnamefont {A.}~\bibnamefont
  {Darras}}, \bibinfo {author} {\bibfnamefont {K.}~\bibnamefont {Peikert}},
  \bibinfo {author} {\bibfnamefont {A.}~\bibnamefont {Rabe}}, \bibinfo {author}
  {\bibfnamefont {F.}~\bibnamefont {Yaya}}, \bibinfo {author} {\bibfnamefont
  {G.}~\bibnamefont {Simionato}}, \bibinfo {author} {\bibfnamefont
  {T.}~\bibnamefont {John}}, \bibinfo {author} {\bibfnamefont {A.~K.}\
  \bibnamefont {Dasanna}}, \bibinfo {author} {\bibfnamefont {S.}~\bibnamefont
  {Buvalyy}}, \bibinfo {author} {\bibfnamefont {J.}~\bibnamefont {Geisel}},
  \bibinfo {author} {\bibfnamefont {A.}~\bibnamefont {Hermann}}, \emph
  {et~al.},\ }\bibfield  {title} {\bibinfo {title} {Acanthocyte sedimentation
  rate as a diagnostic biomarker for neuroacanthocytosis syndromes:
  Experimental evidence and physical justification},\ }\href@noop {} {\bibfield
   {journal} {\bibinfo  {journal} {Cells}\ }\textbf {\bibinfo {volume} {10}},\
  \bibinfo {pages} {788} (\bibinfo {year} {2021}{\natexlab{b}})}\BibitemShut
  {NoStop}%
\bibitem [{\citenamefont {Gelb}\ \emph {et~al.}(2019)\citenamefont {Gelb},
  \citenamefont {Graham}, \citenamefont {Mertz},\ and\ \citenamefont
  {Koenig}}]{gelb2019permeability}%
  \BibitemOpen
  \bibfield  {author} {\bibinfo {author} {\bibfnamefont {L.~D.}\ \bibnamefont
  {Gelb}}, \bibinfo {author} {\bibfnamefont {A.~L.}\ \bibnamefont {Graham}},
  \bibinfo {author} {\bibfnamefont {A.~M.}\ \bibnamefont {Mertz}},\ and\
  \bibinfo {author} {\bibfnamefont {P.~H.}\ \bibnamefont {Koenig}},\ }\bibfield
   {title} {\bibinfo {title} {On the permeability of colloidal gels},\
  }\href@noop {} {\bibfield  {journal} {\bibinfo  {journal} {Physics of
  Fluids}\ }\textbf {\bibinfo {volume} {31}},\ \bibinfo {pages} {021210}
  (\bibinfo {year} {2019})}\BibitemShut {NoStop}%
\end{thebibliography}%

\end{document}


\preprint{APS/123-QED}

\title{Supplemental Material for\\Erythrocyte sedimentation rate :\\influence of binding energy on the rate of soft colloidal gel collapse}

\author{Anil Kumar Dasanna}
\affiliation{Theoretical Physics of Living Matter, Institute of Biological Information Processing and Institute for Advanced Simulation, Forschungszentrum J\"ulich, 52425 J\"ulich, Germany}
\author{Alexis Darras}
\email{alexis.charles.darras@gmail.com}
\affiliation{Experimental Physics, Saarland University, 66123 Saarbruecken, Germany
}
\author{Thomas John}%
\affiliation{Experimental Physics, Saarland University, 66123 Saarbruecken, Germany
}
\author{Gerhard Gompper}
\affiliation{Theoretical Physics of Living Matter, Institute of Biological Information Processing and Institute for Advanced Simulation, Forschungszentrum J\"ulich, 52425 J\"ulich, Germany}
\author{Lars Kaestner}
\affiliation{Experimental Physics, Saarland University, 66123 Saarbruecken, Germany
}
\affiliation{Theoretical Medicine and Biosciences, Saarland University, 66424 Homburg, Germany}
\author{Christian Wagner}
\affiliation{Experimental Physics, Saarland University, 66123 Saarbruecken, Germany
}
\affiliation{Physics and Materials Science Research Unit, University of Luxembourg, Luxembourg City, Luxembourg}
\author{Dmitry A. Fedosov}
\affiliation{Theoretical Physics of Living Matter, Institute of Biological Information Processing and Institute for Advanced Simulation, Forschungszentrum J\"ulich, 52425 J\"ulich, Germany}

\date{\today}

\begin{abstract}
This document gathers Supplemental Figures from the aforementioned article. 
\end{abstract}

\maketitle


\section{Supplemental Figures}
\setcounter{figure}{0}

\setcounter{figure}{0}
\renewcommand{\figurename}{Supplemental Figure}

\begin{figure}[htbp]
    \includegraphics[width=1.\linewidth]{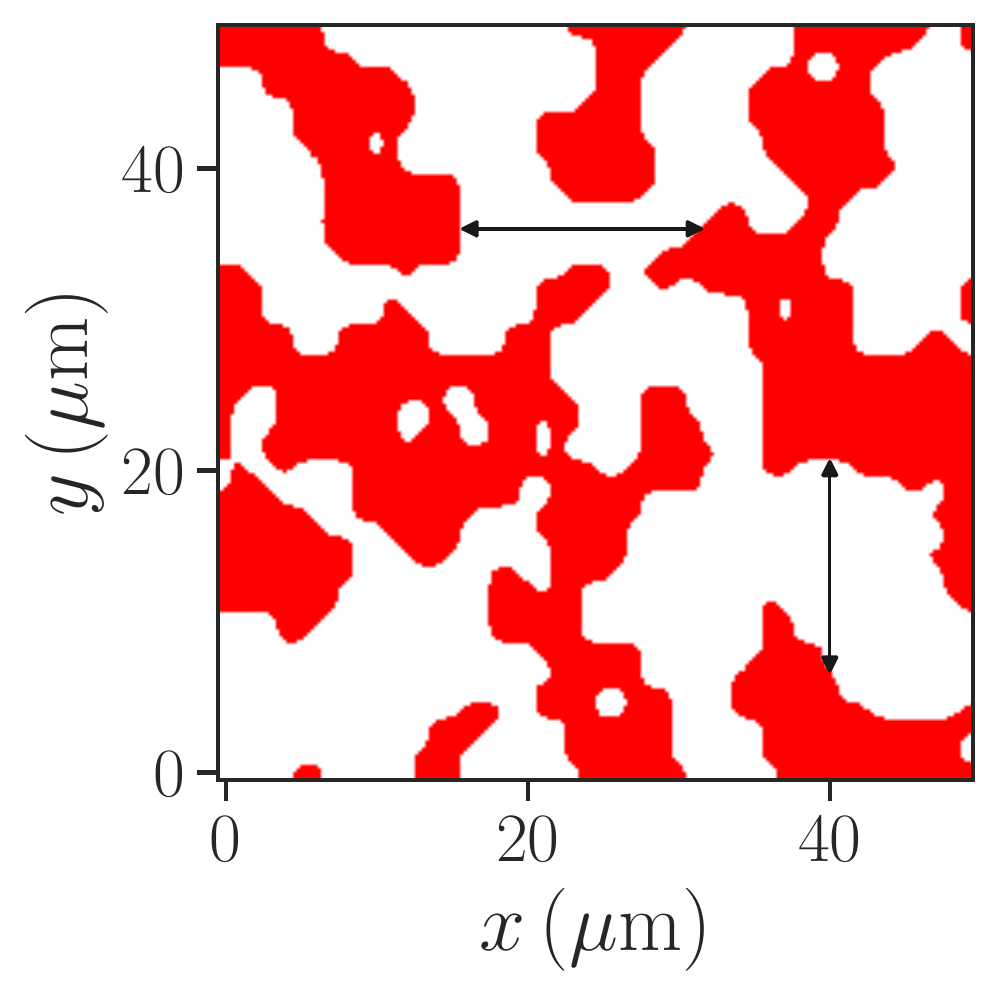}
    \caption{\textbf{Pore sizes determined by numerical simulations.} Example of horizontal $xy$ plane in which the pore size from the simulations is estimated. Every $1\,\mathrm{\upmu}m$ in $x$ and $y$ directions, a point of the slice is identified as being inside (red in the Figure) or outside (white in the figure) the erythrocytes. Then, the lengths of outside distances along $x$ are collected for each $y$ position. The same is done along $y$. The distribution of such lengths is shown in Supplemental Figure \ref{SuppFig4}, and the average value is considered to be the characteristic pore size in Figure 6. This $xy$ slide has been obtained with a hematocrit of $\phi=0.35$ and an aggregation strength $\epsilon=2.5~\kbt$.}
    \label{SuppFig3}
\end{figure}

\begin{figure}[htbp]
    \includegraphics[width=1.\linewidth]{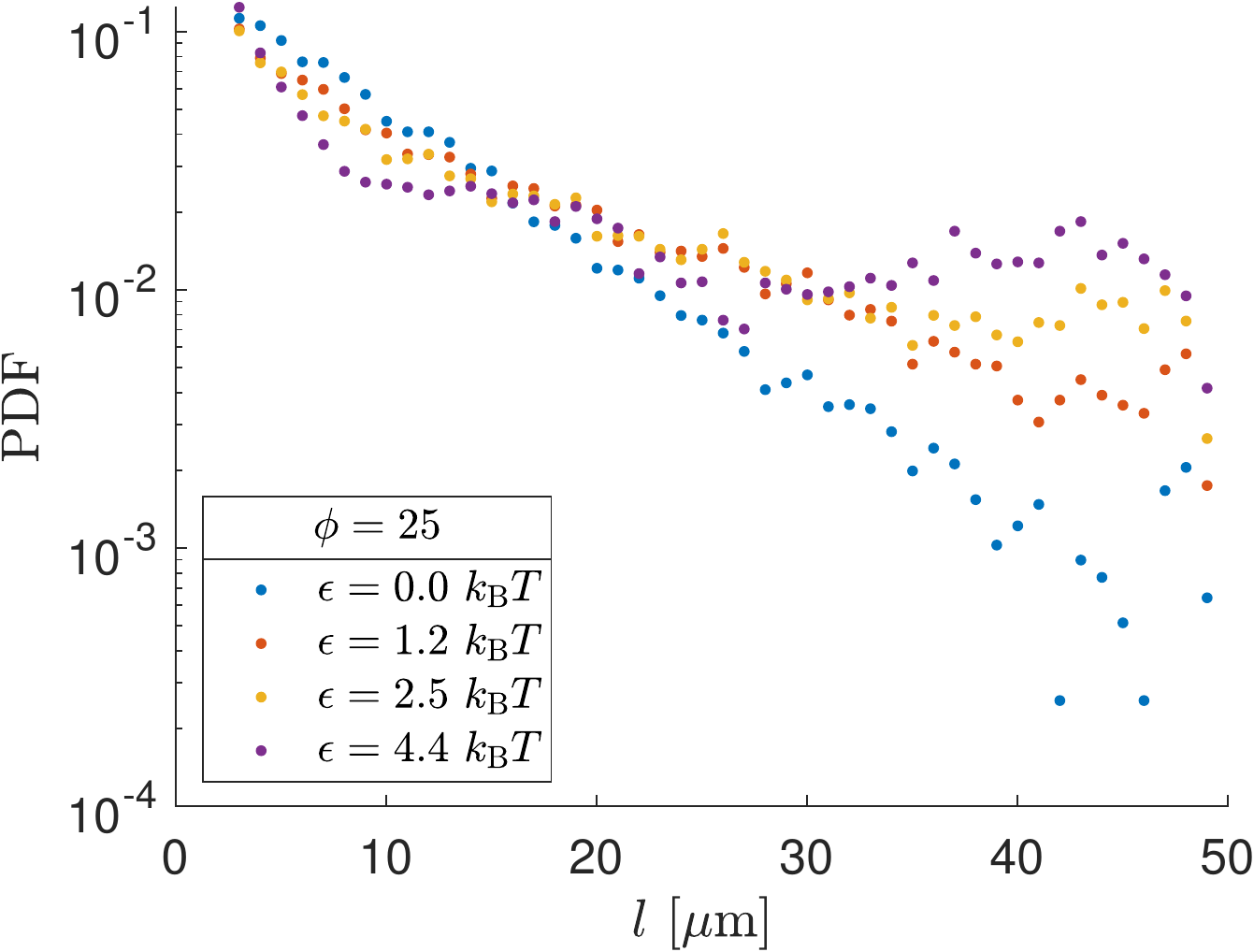}
    \caption{\textbf{Holes size PDF.} The probability density functions for characteristic holes sizes $l$ in the percolating network are shown for various interaction energies $\epsilon$. Increasing the interaction energy leads to an increase in the probability of holes with a larger size. Data are shown for hematocrit $\phi=0.25$.
}
    \label{SuppFig4}
\end{figure}

\newpage
~~~
\newpage
\section{Legends of Supplemental Movies and Supplemental Figures}
\setcounter{figure}{0}
\renewcommand{\figurename}{Supplemental Movie}
\renewcommand{\thefigure}{S\arabic{figure}}

As long as the manuscript is not published in a peer-reviewed journal, the Supplemental Movie can be accessed at \url{https://drive.google.com/file/d/1G1DhxPomMr69I5UfnlPMP9_544fK7dUg/view?usp=sharing}.

\begin{figure}[!htbp]
    \caption{\textbf{Illustrative movie generated from numerical simulations} (H40\_ImageJ.avi) Numerical simulation of the sedimentation of erythrocytes, with hematocrit of $\phi=0.4$. The movie shows a vertical slice of the simulation domain, with a depth of view of $10\,\mathrm{\upmu}m$. One can see the formation of a percolating network of erythrocytes falling down, after a transient period where some cells are advected upward by the flow of the liquid. The movie has a duration of $140\,\mathrm{s}$ in real time units.
}
    \label{SuppMov3}
\end{figure}